\documentclass[twocolumn,showpacs,preprintnumbers,prl,superscriptaddress]{revtex4-2}
\usepackage{graphicx,color}
\usepackage{float}
\usepackage{amsmath}
\usepackage{bm}
\usepackage{tkz-berge,tikz}
\usepackage[qm,braket]{qcircuit}

\usepackage[pdftex,colorlinks=true, linkcolor = blue, citecolor=blue,urlcolor=blue, bookmarksnumbered=true, bookmarksopen=true]{hyperref}
\usepackage{appendix}
\usepackage{soul}

\usepackage[normalem]{ulem}

\begin{document}

\title{Projection algorithm for state preparation on quantum computers}

\author{I. Stetcu}
\affiliation{
  Los Alamos National Laboratory, Theoretical Division, Los Alamos, New Mexico 87545, USA}
\author{A. Baroni}
\affiliation{
  Los Alamos National Laboratory, Theoretical Division, Los Alamos, New Mexico 87545, USA}
\affiliation{National Center for Computational Sciences, Oak Ridge National Laboratory, TN 37831, USA}
\author{J. Carlson}
\affiliation{
  Los Alamos National Laboratory, Theoretical Division, Los Alamos, New Mexico 87545, USA}
  
\date{\today}
\preprint{LA-UR-22-30149 Version 3}

\begin{abstract}
    We present an efficient method to prepare states of a many-body system on quantum hardware, first isolating individual quantum numbers   and then using time evolution  to isolate the energy. Our method in its simplest form requires only one additional auxiliary qubit. The total time evolved for an accurate
    solution is proportional to the ratio of the spectrum range of the trial state to the gap to the lowest excited state, a substantial improvement over other projection algorithms, and the accuracy increases exponentially with the time evolved. Isolating the quantum numbers is efficient  because of the known eigenvalues, and  increases the gap thus shortening  the propagation time required. The success rate of the algorithm, or the probability of producing the desired state, is a simple function of measurement times and phases and is dominated by the  square overlap of the original state to the desired state. We present examples from the nuclear shell model and the Heisenberg model.
    We compare this algorithm to previous algorithms for short evolution times and discuss potential further improvements.
\end{abstract}

\maketitle


Quantum computers are expected to have an exponential advantage in calculating the dynamics of strongly-interacting quantum systems.  This has many possible applications in all fields of physics, from neutron scattering of materials to response functions studied experimentally in cold atom systems to high-energy neutrino electron and neutrino scattering from nuclei to hadronization in QCD.  In each of these applications, though, one must often prepare a specific initial state.  State preparation is a
critical initial step in simulating quantum dynamics
on many-body systems.

While preparing the ground state of general Hamiltonians is in principle QMA-complete~\cite{Kempe:2004}, in practice accurate
ground states can be prepared on classical computers using
approximate algorithms including Quantum Monte Carlo~\cite{Foulkes:2001,Carlson:2015}, Tensor Networks~\cite{Orus:2019}, and Coupled Cluster~\cite{Bartlett:1981,Hagen2014} methods.
On a quantum computer it has been shown that a ground state
can be prepared if the ground state energy, spectral gap
to a low-lying excitation, and total width of the spectrum
are known and if we can prepare an initial state with
a non-zero overlap with the desired eigenstate \cite{Ge2018}.

In this paper we present an efficient Quantum Projection Filter
(QPF) algorithm that requires
shorter evolution times, particularly useful 
for early fault tolerant quantum computers.  We plan to investigate
applications to advanced NISQ devices with small noise in a future work.
We optimize the measurements required to produce the ground state assuming minimal knowledge of the spectra of the initial state.
We assume the dominant limitation on near-term devices
is the total evolution time $t$ required to obtain
$\exp ( - i H t) $
in the physical space, where $H$ is the Hamiltonian operator.
This is a fundamental limitation for a general exact algorithm due to the no fast-fowarding theorem \cite{Berry_2006}.

We demonstrate that accurate states can be prepared
with a total propagation time of a small multiple of the inverse
spectral gap. By projecting on good quantum numbers, 
we increase the relevant spectral
gap, thus reducing the evolution time required.  We then
perform energy projection assuming minimal knowledge
of the ground state energy, the spectrum width, the increased gap, 
and the overlap, here optimizing
both measurement times and phases. The accuracy of this method increases exponentially with time, so improvements in hardware while reducing errors will have a large impact. This algorithm takes advantage
of good initial variational states which shorten the required evolution time by reducing  the highest energy states having overlap with the initial state.

Several general classes of algorithms for initial state preparation on quantum computers have been presented in the literature.  Variational algorithms like VQE~\cite{Kandala:2017HE,Tilly:2021} often provide a critical first step in state
preparation, but while these methods are variational,
providing an upper bound to the ground state energy,
they are not exact and not always systematically improvable
with known convergence rates.

Exact algorithms for initial state preparation
similarly can be divided into several general categories.  Adiabatic evolution can be used to transform from a known ground state of a different
Hamiltonian to the ground state of the desired Hamiltonian by a slowly-evolving time-dependent
Hamiltonian. This method is exact but requires
long circuit depths and relies upon the adiabatic
evolution of states with no sharp avoided level 
crossings~\cite{Farhi2000,Albash2018}. 
Projection methods have also been developed to reduce Trotter errors \cite{PhysRevLett.112.120406,*PRXQuantum.2.010323,*PRXQuantum.2.040311,*PRXQuantum.3.020324,*Halimeh_2022} and for symmetry restoration \cite{PhysRevLett.125.230502,*PhysRevA.104.062435,*PhysRevC.105.024324}. 
Some of these algorithms mimic classical 
computations,
like quantum imaginary time evolution (QITE) and quantum Lanczos (QLANCZOS) \cite{Motta:2020,PhysRevResearch.4.033121,Turro2022,Jouzdani2022}, which are quantum-computer analogues of imaginary-time evolution employed in quantum Monte Carlo and the Lanczos algorithm. These methods are extremely successful
on classical computers with applications in 
cold and hot dense matter \cite{Holmes2022quantumalgorithms}, atomic and molecular
systems, nuclear physics, and QCD. They often suffer
from sign problems but very accurate approximations
to the path integral have been developed. Quantum
computers by nature provide unitary evolution, and
QITE and QLANCZOS emulate the classical computing
equivalent through hybrid classical-quantum approaches. These hybrid algorithms suffer
an exponential overhead in general due to the
classical optimization procedure.

Projection methods based directly upon unitary evolution have also been studied \cite{Ge2018,Dong2022,Keen2021,Choi2021}. 
We propose a similar but more flexible algorithm for
state preparation that for the first time takes into account the symmetries exhibited by the problems considered. Our algorithm first projects
out the desired quantum numbers of the state to
increase the relevant spectral gaps 
and then uses unitary time evolution to filter 
the system by energy. Similar to the other 
algorithms, it is a probabilistic algorithm
in that the state will be prepared successfully
with a probability determined by the squared overlap of the
initial state with the desired state and the
times and phases used in the measurements. 

The total
evolution time required, a key indicator of the
efficiency of the algorithm, is proportional to the
inverse of the ratio of the gap of the desired state to nearby states with the same quantum numbers to
the total width of the initial state energy distribution. The algorithm can be used to prepare states with various
quantum numbers and energies. 

We describe a QPF algorithm with one auxiliary qubit,
similar to the algorithms for both eigenvector preparation of Refs.~\cite{Ge2018,Choi2021} and response function calculation of Ref.~\cite{Rogerro2019}. Extensions
to multiple auxiliary qubits should be valuable given
sufficient quantum resources and sufficiently low error
rates. 

 Attaching an ancilla qubit $a$ in a 0 state so that the entangled 
system state is $|\psi \rangle \otimes |0\rangle_a$, we perform a series of unitary transformations $\exp \left(-i (t_i O + \delta_i) \otimes Y_a\right)$, so that for each iteration the new state of the system becomes
\begin{equation}
    |\psi(t)\rangle  =  \cos(t_i O+\delta_i)  |\psi\rangle \otimes |0\rangle_a 
    + \sin(t_i O+\delta_i)|\psi\rangle \otimes |1\rangle_a.
\label{eq:timeevolve}
\end{equation}
After each transformation, we measure qubit $a$, which, for selected times $t_i$ and phases $\delta_i$, projects on the desired physical subspace. A schematic circuit is shown in Eq.~(\ref{eq:circuit}).
The operator $O$ defines the relevant quantum numbers of the system 
(e.g. $ J^2$, particle number or total momentum) 
or the Hamiltonian $H$ for energy projection. The probability of reading off qubit $a$ in state $|0\rangle$ is $\langle \psi|\cos^2(t_i O+\delta_i)|\psi\rangle$ and the success probability of the algorithm will be given by the product of individual probabilities after all measurements~\cite{supplement}.
\begin{equation}
    \Qcircuit @C=1.25em @R=1.1em @!R{
        \lstick{\ket{0}_a}&\qw 
        &\multigate{1}{\ e^{-iY_a\otimes(t_iO+\delta_i)} \ }&\meter&\qw\\
        \lstick{\ket{\psi}}&\qw&\ghost{\ e^{-iY_a\otimes(t_iO+\delta_i)} \ }&\qw&\qw
        }
        \label{eq:circuit}
\end{equation}
We choose the number of different measurements that scales logarithmically with the total time (sum of all individual  times). 
Trotterization of each evolution or qubitization
may require significant resources per measurement
related to the time evolution required. 


We perform a series of time evolutions followed by measurements. Requiring each time the qubit $a$ to be in state $0$ yields
a physical state 
$| \psi \left(t = \sum_i t_i\right)  \rangle  = 
{\cal N} \prod_i \cos (t_i O + \delta_i) |\psi (t=0)\rangle,$
with a probability of success (all 0 measurements)
given by the product of the squared cosines. Below we
call the product of phase shifted cosines the filter function $f  = \prod_i \sum_\alpha \cos(t_i o_\alpha + \delta_i)$, where $o_\alpha$ are the eigenvalues of $O$.

Projecting on quantum numbers is a crucial initial step, and is readily accomplished
since the spectrum is known in advance and is very regular. For symmetry projections, we take all $\delta_i=0$. Assume for the moment that we desire a spin zero (ground) state. 
Choosing $t_1 =  \pi /4$ removes concurrently $J=1, 2, 5, 6, 9, 10, 13, 14,\ldots $ values, as $\cos ( J(J+1)\pi/4)=0$. 
To project out other $J$ states requires more
iterations.  Taking
 $t_{i+1} = t_i / 2$, 
one projects out successfully larger values of
$J$, including values missed in previous iteration(s), with shorter and shorter 
additional time evolutions. The initial (large) value of $t_1$ defines the
resolution of the projection, and subsequent $t_i$
with successfully smaller values  gives a projection function value of zero for larger and larger values of
$J$.  The largest zero value for the projection
function increases exponentially with the number
of measurements.  Thus, with this choice of times
(and all zero phases) the filter function rapidly
approaches a delta function centered on zero.

Similar sequences work well for other dispersion
relations: linear as for particle number or
quadratic like a typical non-relativistic dispersion
relation. In each case the longest time
is governed by the lowest excitation, 
with additional exponentially shorter times. The
total time evolution approaches $2 t_1$ after
all steps.  Different eigenvalues can be projected
out by shifting the operator by a constant.
The total success probability to project onto the targeted space is the square of the overlap between the initial state
and the desired quantum number for zero phases. The final state
after quantum number projection retains the space spanned by the
eigenstates with the same quantum number
and the same relative amplitudes as in the
original state.  Often the number of energy eigenstates is
reduced dramatically in this step, and the
corresponding gaps similarly increased, which
increases the efficiency of subsequent energy projection.


The energy can be projected similarly. If the spectrum is known from classical computations, we can apply the same algorithm described above for symmetry projection to prepare the initial state by removing individual undesired states. However, the method can be used with limited knowledge of the spectrum, assuming that the desired energy, the approximate width of the spectrum and the approximate gap to nearby states are known to good enough precision ($10-20\%$). If the spectrum is not known, one can obtain information about the eigenvalues and the overlap with the eigenvectors of the Hamiltonian by running the circuit in Fig. 12 of Ref. \cite{Soma-SpectrumState}. This will also reduce the evolution time required for a well-chosen variational trial state, as high energy states will not contribute. In the remaining discussion
we assume the Hamiltonian is shifted and scaled to have
a range of eigenvalues between zero and one. We
present ground state projection where the desired
state is at zero energy. Excited states can be prepared similarly.

For the energy the starting state typically
does not have known spectra and overlaps.
Using only the minimal information that the
shifted ground state is near zero energy,
an estimate of
the gap  $\Delta_e$, and a rough estimate of the overlap
between the trial and desired state $\alpha_e$, we can design a filter that
quickly reaches the ground state.
To optimize the times and phases in the filter, we assume a pseudo-Hamiltonian $\tilde H$ with a zero energy ground state,
a gap $\Delta_e$
and a uniform spectra from the gap $\Delta_e$ 
to 1 (maximum dimensionless energy in the spectrum) with small random perturbations to the
eigenvalues to 
preclude artificial minima for regular spectra.
We assume the initial spectral distribution with an amplitude governed by $\alpha_e$ at zero energy and equal amplitudes on each 
excited state of $\tilde H$.
We then apply the equivalent of the quantum filtering algorithm to this state and optimize the times $t_i$ and phases $\delta_i$ to maximize the squared overlap of the final state with the ground state. This optimization consists in retaining $\cos(\tilde Ht_i+\delta_i)|\psi(t_{i-1})\rangle$ after each time step (note that for optimization all states are written as superposition of $\tilde H$ eigenstates), and can be done beforehand
on a classical computer since the number
of measurements (parameters) is logarithmic
in the desired accuracy of the output state.

\begin{figure}
\includegraphics[scale=0.4]{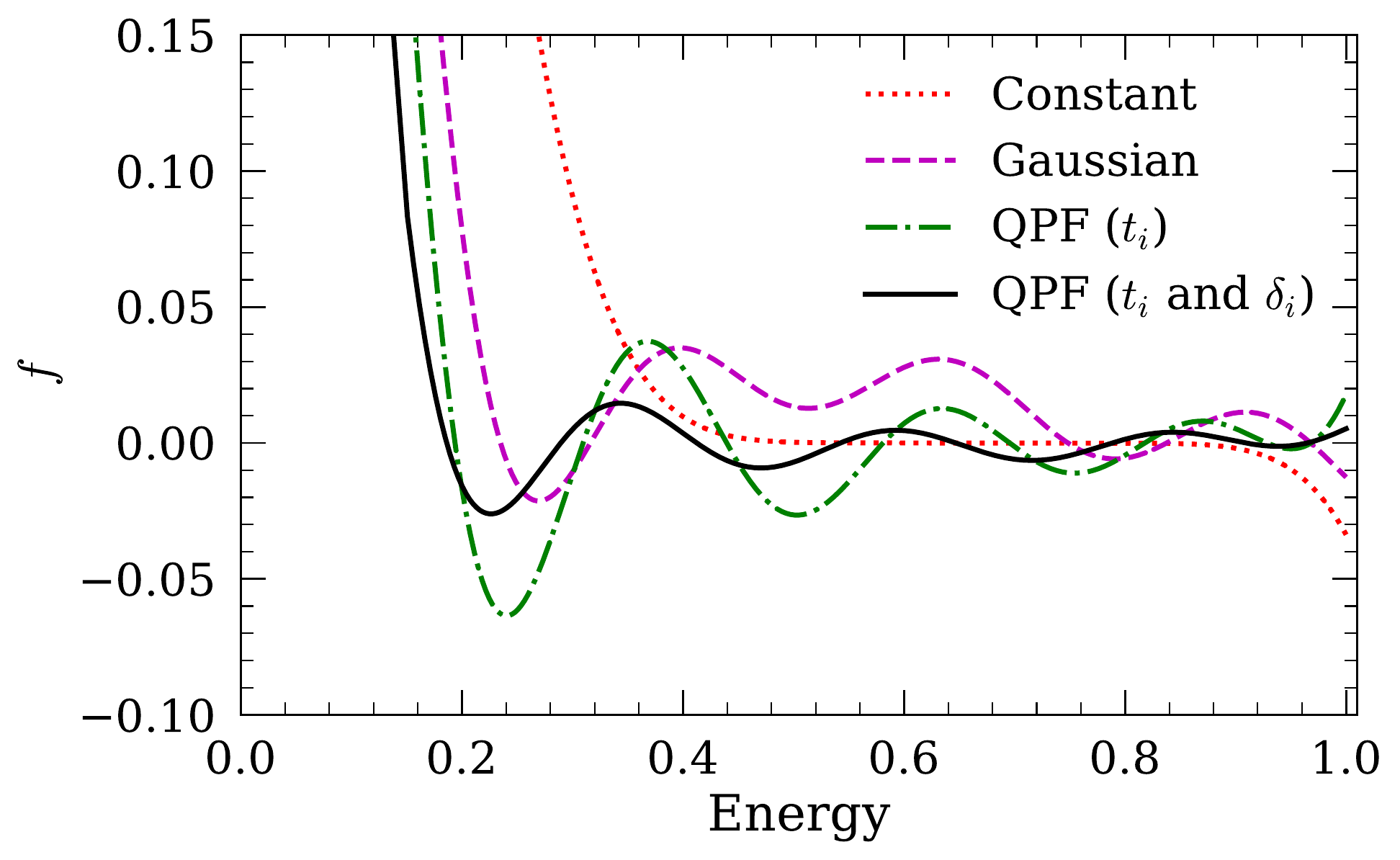}
\caption{Optimized filter function 
compared with other time evolution choices. The energy is rescaled from 0 to 1, with the targeted state at 0 and an assumed gap of 0.15. Constant times are similar to the \citet{Ge2018} algorithm and Gaussian-sampled times to the Rodeo algorithm \cite{Choi2021}. For an example of a filter for the $J^2$ operator, see Fig. 1 in the Supplemental Material \cite{supplement}.}
\label{fig:filter}
\end{figure}

In Fig. \ref{fig:filter} we show the resulting
filter function for $\Delta_e=0.15$ and an
assumed  ground state overlap of 0.05. The filter
function ($\delta_i=0$) for constant, Gaussian, and exponential time distributions are all shown, in each case limiting the total time evolution
to $\pi/\Delta_e$. The filter function
is larger for the ground state compared to other excitations, particularly low-lying excitations,
guaranteeing exponential convergence when iterated. The optimization improves the
filtering of low-energy excitations in particular.
Non-zero phase shifts allow the filter function to be linear and decreasing near the ground state,
and enable a total projection linear in the inverse gap rather than quadratic. We show an example where the ratio of the gap to the full spectral width is $10^{-4}$ in 
\cite{supplement}.

Upon convergence,
the cumulative success probability for obtaining
 the ground state (or a very good approximation) starting from the initial trial state $|\psi(t=0)\rangle$ is given by
$P_\mathrm{success} = f(0)^2|\langle \Psi_0 | \psi(t=0) \rangle|^2$,
where $\Psi_0$
is the desired ground state and $f(0)=\prod_i \cos\delta_i$.
The factor $f(0)^2$ is one for cases where the phases
are set to zero. Otherwise it is an output of the
optimization but typically close to one. The function to be optimized
can be made closer to one by increasing $\alpha_e$, with a potentially
slightly longer evolution time required for convergence.
The success probability can also be reduced slightly
if the exact ground state energy is unknown.
The success probability is the product
of success probabilities for each measurement,
each of which is less than one, but the overall success
is finite because the state is improving after
each measurement, with individual measurement
success probabilities rapidly approaching one \cite{supplement}.

The times $t_i$ and the phase shifts $\delta_i$
can be optimized based upon facts known about the
spectra and the overlap of the initial state with
eigenstates of different energies. 
For example if there are low lying quasiparticles with small widths
and well defined energies, the algorithm could be adjusted to filter them out quickly.  Phase shifts could be useful to help project out isolated low-lying states. Non-zero phases give asymmetry between positive and negative energies and could help remove isolated low-lying states where the largest evolution 
time required by zero phase shifts is very large.
Further details are given in the Supplemental Material \cite{supplement}.

The times and phases can also potentially be optimized 
to partially take into account
statistical or systematic errors in the quantum hardware.  For example errors may increase with time
due to noise in the system and potentially due to
Trotter errors.
This will limit the possible energy resolution,
but in principle the times and phases could be further optimized given a model of the errors.




We first show results for a simple  Heisenberg model
\begin{equation}
H = \sum_{<i,j>} \sigma_i \cdot \sigma_j + \sum_i h \sigma_i \cdot {\hat z},
\end{equation}
where the sum over $i$ and $j$ runs over nearest neighbors,
the $\sigma$ are Pauli SU(2) spinors and $h$ represents
a coupling to an external magnetic field.
On larger lattices this problem has been extensively studied in the
literature, using exact diagonalization, Quantum Monte Carlo, and other methods. The ground state has
been calculated extremely accurately using QMC
methods, which for a bipartite lattice does not suffer
from a sign problem. The ground state is $J=0$
for $h=0$ but can have a large $J$ for large magnetic
couplings $h$.
Here we consider a 2D 4x4 square lattice in periodic boundary conditions. For simplicity we take $h=0$,
for finite $h$ we would have just an overall energy shift for each $J_z$ subspace.

Both the total spin and the total third component
of spin are good quantum numbers for this Hamiltonian.
In total there are $2^{16} = 65,536$ many-body states.
12,870 states have total $J_z = 0$, or an
equal number of up and down spins. In principle these
could be isolated from a more general state
by projecting upon $J_z$. Here we assume the
initial state has good $J_z =0$.

For the initial  state  we use 
the Neel state with all spins up (down) on even (odd) sublattices.
This state has good $J_z=0$ but  not 
good $J^2$. The initial state has a fairly large $\langle J^2 \rangle$ and not a very accurate energy
(scaled energy $\approx 0.17$)
The maximum possible $J$ in this system is $J=N/2 = 8$.
To project out all the $J>0$ states requires
three iterations, the last one would be sufficient to
project out even higher $J$ on a bigger lattice.
For this case the projection exactly projects
out eigenstates with $J>0$ resulting in a state
with $\langle J^2 \rangle = 0$. The energy is
lowered to $E\approx0.06$ after the $J^2$ projection.


The success probability of the $J=0$ projection is approximately 11\%, corresponding to the fraction of the initial state probability in the $J=0$ subspace. This could be improved with better trial states, but would be smaller for larger lattices.
The gap from the ground state to the lowest relevant
excited state is significantly enhanced with
$J^2$ projection, from
approximately 0.03 
to 0.15
of the total spectrum width.

\begin{figure}
\includegraphics[scale=0.5]{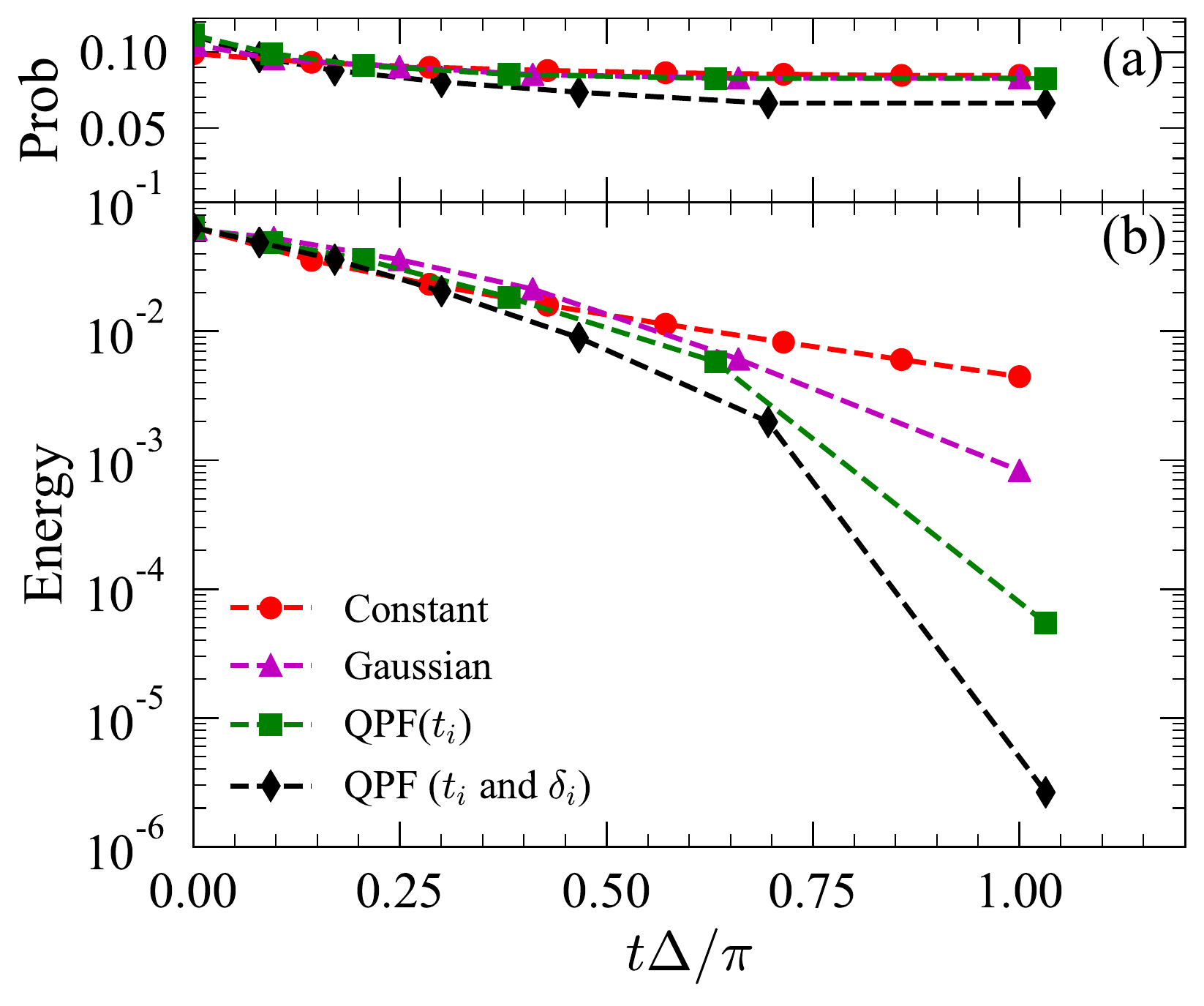}
\caption{Success probabilities (a) and energies (b) as a function of time for different projection algorithms after $J^2$
projection for the Heisenberg model. Success of the $J^2$ projection
is at approximately 0.11 which is the $t=0$ starting
point for overall success. The target energy is at 0.}
\label{fig:evstauh}
\end{figure}

We then do an energy projection after obtaining the large gap between the ground and first excited state.
We compare the energy as a function of time for constant, Gaussian, and optimized times without and with phases in the bottom panel of Fig.~\ref{fig:evstauh}. 

The plot compares results for total propagation time of $t = \pi$/$\Delta$. For the constant and Gaussian cases
we choose the number of steps to optimize the energy with
a fixed total propagation time.
For the Gaussian case we average over choices of 7 steps from a normal distribution and scaled them to a total evolution time
of $t = \pi/\Delta$. For the QPF  case with times and
phases we
chose times and phases assuming a gap of 0.15 and a
5\% initial ground state probability. 
We find six time and phase measurements gives the
optimum minimization, adding further
measurements does not improve the optimization unless more
is known about the initial energy distribution.
The small non-zero phases found slightly reduce the success probability (see top panel in Fig.~\ref{fig:evstauh}), but increase the accuracy of the prepared ground state.
Similarly we show results with optimized times $t_i$ but zero phases.

A \textsc{qiskit} implementation of the 4 by 4 lattice for the Heisenberg problem uses 21.8k \textsc{cnot} gates and 23k \textsc{rz} and \textsc{ry} gates, producing in both cases an approximation to the exact ground state having an energy with a relative error compared to the exact energy of $0.6$\% . Using the results of Ref.~\cite{Kliuchnikov2016,Gheorghiu2022} the number of T gates needed to synthesize the above number of \textsc{rz} with precision $10^{-7}$ is approximately $1.5\times 10^6$.

Finally, we have considered the shell model Hamiltonian as in previous publications \cite{Stetcu2022var,PhysRevC.106.034325}, in order to test QPF for a problem with the complexity of the nuclear many-body system. This is
similar to a lattice model in that there is a finite
single-particle basis, but the interactions generally span the space with only angular momentum conservation as a symmetry. Core states are regarded as fully occupied, and the interaction between active nucleons is phenomenologically adjusted to reproduce energy levels in several nuclei. In this paper, we show results for $^{10}$B, i.e., three protons and three neutrons active in the $p$ shell interacting via the phenomenological Hamiltonian \cite{COHEN19651}, assuming $^{4}$He inert core. The spin of the ground state in this case is 3, and in this case the operator used to project on states with $J=3$ is $J^2-12$. Other examples for even-even and odd-mass nuclei are given in the Supplemental Material \cite{supplement}.

In Fig. \ref{fig:b10} we compare the improvement in the ground state energy for times up to $t=\Delta/\pi\simeq 1.2$, where $\Delta$ the exact excitation energy of the first $3^+$ excited state. The success probabilities shown in \ref{fig:b10}(a) as a function of time converge to $|\langle \Psi_0|\psi(t=0)\rangle|^2$ in the absence of algorithmic and hardware errors. We start from the same state, and apply QPF using the energy filter only (QPFE), or after we project on the space with $J=3$ components only, see Fig. \ref{fig:b10}(b). All the convergence patterns are similar, with only the constant time being significantly slower than the others. The main advantage of QPF, which first projects out states with $J\neq3$ in this case and uses optimized time steps and phases, is shown with filled diamonds in \ref{fig:b10}(b).  The final state can be made more accurate by
iterating with the same times 
or by optimizing times and phases for a longer
total evolution time.  In either case the
state converges exponentially.

\begin{figure}[tb]
    \includegraphics[scale=0.5]{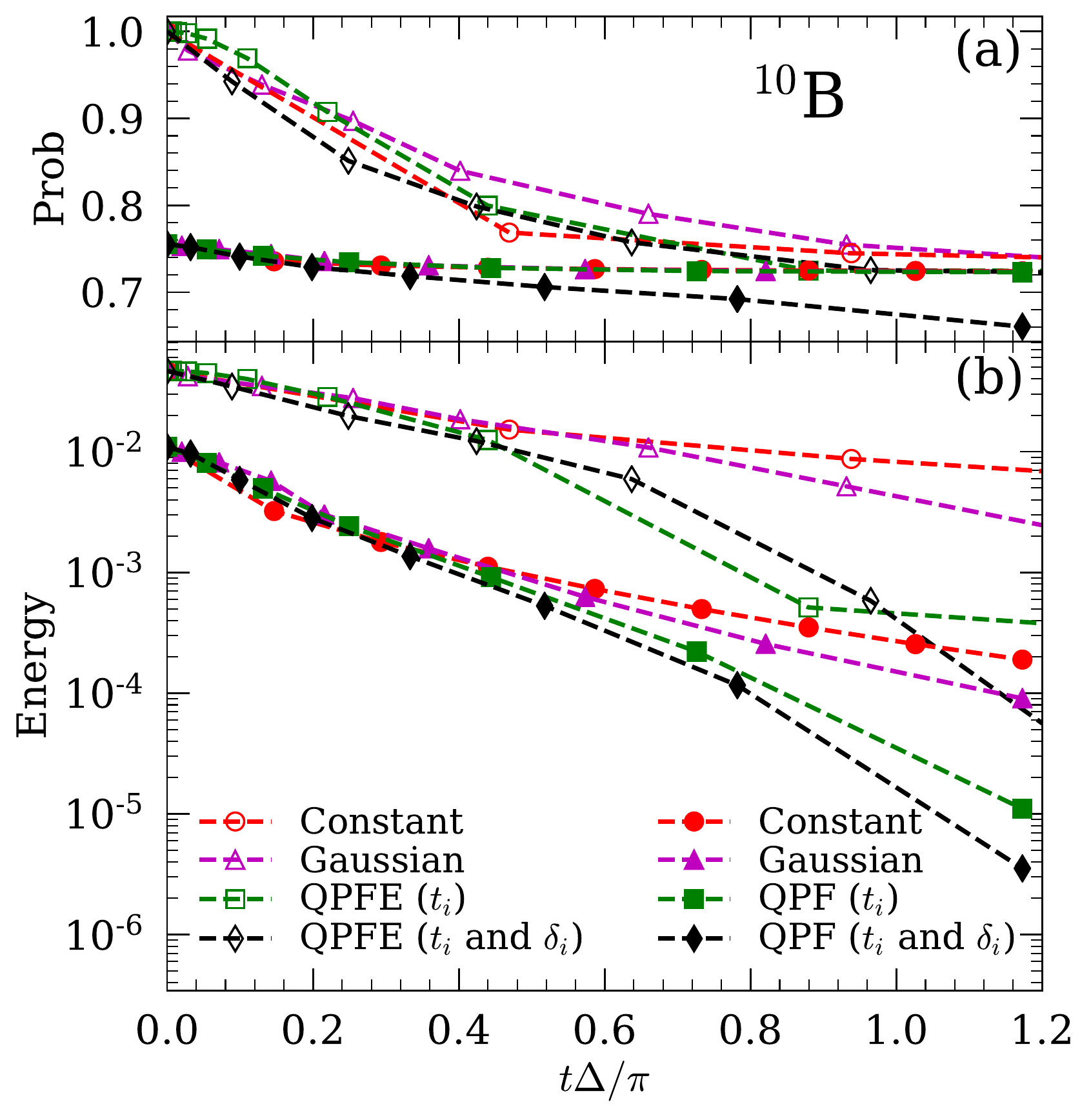}
    \caption{The success probabilities (a) and energies (b) as a function of time for preparing the ground state for $^{10}$B in the $p$ shell model space, starting from the HF solution. We show constant time (red circles), Gaussian-sampled times (purple triangles), QPFE (green squares), and QPFE with optimized times and phases (black diamonds), and with the same filed symbols the Constant, Gaussian-sampled times, QPF and QPF with optimized times and phases, but after the trial HF state has been projected on $J=3$. The exact targeted energy is at zero.}
    \label{fig:b10}
\end{figure}


The algorithm presented here
is very efficient in terms of number of auxiliary qubits, total projection time, and success probability. It can be further optimized given
additional information on the initial state overlaps as a function of quantum numbers and spectra, and to some degree given noisy hardware. It may also
be possible to further improve the efficiency
by measuring the frequency of auxiliary qubit
patterns not contributing to the ground state.
The success probability may be increased by
considering alternative dynamics.
 
We further expect that very similar algorithms will be extremely useful for linear response studies,
particularly when one is interested in the response for specific energies,
an important early application of quantum computers in various fields.  In addition they may find useful applications in explicit final state measurements of time-evolved quantum systems.\\

\section*{Acknowledgments}


We thank C.~W. Johnson and R. Weiss for the feedback on the manuscript.
 AB thanks Y. Subasi, A. Roggero, E. Dumitrescu, Y. Wang, T. Morris for discussions regarding the LCU-like algorithms for state preparation.
 This work was carried out under the auspices of the National Nuclear Security Administration of the U.S. Department of Energy at Los Alamos National Laboratory under Contract No. 89233218CNA000001, and Oak Ridge Leadership Computing Facility at the Oak Ridge National Laboratory, which is supported by the office of Science of the U. S. Department of Energy under contract No. DE-AC05-00OR22725. IS and JC gratefully acknowledge  partial support by the Advanced Simulation and Computing (ASC) Program. AB’s work is supported by the U.S. Department of Energy, Office of Science, Nuclear Physics Quantum Horizons initiative. This work was partially funded by the U. S. Department of Energy, Office of Science, Advanced Scientific Computing Program Office under FWP ERKJ382. 
 JC and AB also acknowledge the Quantum Science Center
 for partial support of their work on this project.

\bibliography{references}

\begin{thebibliography}{36}%
\makeatletter
\providecommand \@ifxundefined [1]{%
 \@ifx{#1\undefined}
}%
\providecommand \@ifnum [1]{%
 \ifnum #1\expandafter \@firstoftwo
 \else \expandafter \@secondoftwo
 \fi
}%
\providecommand \@ifx [1]{%
 \ifx #1\expandafter \@firstoftwo
 \else \expandafter \@secondoftwo
 \fi
}%
\providecommand \natexlab [1]{#1}%
\providecommand \enquote  [1]{``#1''}%
\providecommand \bibnamefont  [1]{#1}%
\providecommand \bibfnamefont [1]{#1}%
\providecommand \citenamefont [1]{#1}%
\providecommand \href@noop [0]{\@secondoftwo}%
\providecommand \href [0]{\begingroup \@sanitize@url \@href}%
\providecommand \@href[1]{\@@startlink{#1}\@@href}%
\providecommand \@@href[1]{\endgroup#1\@@endlink}%
\providecommand \@sanitize@url [0]{\catcode `\\12\catcode `\$12\catcode
  `\&12\catcode `\#12\catcode `\^12\catcode `\_12\catcode `\%12\relax}%
\providecommand \@@startlink[1]{}%
\providecommand \@@endlink[0]{}%
\providecommand \url  [0]{\begingroup\@sanitize@url \@url }%
\providecommand \@url [1]{\endgroup\@href {#1}{\urlprefix }}%
\providecommand \urlprefix  [0]{URL }%
\providecommand \Eprint [0]{\href }%
\providecommand \doibase [0]{https://doi.org/}%
\providecommand \selectlanguage [0]{\@gobble}%
\providecommand \bibinfo  [0]{\@secondoftwo}%
\providecommand \bibfield  [0]{\@secondoftwo}%
\providecommand \translation [1]{[#1]}%
\providecommand \BibitemOpen [0]{}%
\providecommand \bibitemStop [0]{}%
\providecommand \bibitemNoStop [0]{.\EOS\space}%
\providecommand \EOS [0]{\spacefactor3000\relax}%
\providecommand \BibitemShut  [1]{\csname bibitem#1\endcsname}%
\let\auto@bib@innerbib\@empty
\bibitem [{\citenamefont {{Kempe}}\ \emph {et~al.}(2004)\citenamefont
  {{Kempe}}, \citenamefont {{Kitaev}},\ and\ \citenamefont
  {{Regev}}}]{Kempe:2004}%
  \BibitemOpen
  \bibfield  {author} {\bibinfo {author} {\bibfnamefont {J.}~\bibnamefont
  {{Kempe}}}, \bibinfo {author} {\bibfnamefont {A.}~\bibnamefont {{Kitaev}}},\
  and\ \bibinfo {author} {\bibfnamefont {O.}~\bibnamefont {{Regev}}},\
  }\bibfield  {title} {\bibinfo {title} {{The Complexity of the Local
  Hamiltonian Problem}},\ }\href@noop {} {\bibfield  {journal} {\bibinfo
  {journal} {arXiv e-prints}\ ,\ \bibinfo {eid} {quant-ph/0406180}} (\bibinfo
  {year} {2004})},\ \Eprint {https://arxiv.org/abs/quant-ph/0406180}
  {arXiv:quant-ph/0406180 [quant-ph]} \BibitemShut {NoStop}%
\bibitem [{\citenamefont {{Foulkes}}\ \emph {et~al.}(2001)\citenamefont
  {{Foulkes}}, \citenamefont {{Mitas}}, \citenamefont {{Needs}},\ and\
  \citenamefont {{Rajagopal}}}]{Foulkes:2001}%
  \BibitemOpen
  \bibfield  {author} {\bibinfo {author} {\bibfnamefont {W.~M.}\ \bibnamefont
  {{Foulkes}}}, \bibinfo {author} {\bibfnamefont {L.}~\bibnamefont {{Mitas}}},
  \bibinfo {author} {\bibfnamefont {R.~J.}\ \bibnamefont {{Needs}}},\ and\
  \bibinfo {author} {\bibfnamefont {G.}~\bibnamefont {{Rajagopal}}},\
  }\bibfield  {title} {\bibinfo {title} {{Quantum Monte Carlo simulations of
  solids}},\ }\href {https://doi.org/10.1103/RevModPhys.73.33} {\bibfield
  {journal} {\bibinfo  {journal} {Reviews of Modern Physics}\ }\textbf
  {\bibinfo {volume} {73}},\ \bibinfo {pages} {33} (\bibinfo {year}
  {2001})}\BibitemShut {NoStop}%
\bibitem [{\citenamefont {Carlson}\ \emph {et~al.}(2015)\citenamefont
  {Carlson}, \citenamefont {Gandolfi}, \citenamefont {Pederiva}, \citenamefont
  {Pieper}, \citenamefont {Schiavilla}, \citenamefont {Schmidt},\ and\
  \citenamefont {Wiringa}}]{Carlson:2015}%
  \BibitemOpen
  \bibfield  {author} {\bibinfo {author} {\bibfnamefont {J.}~\bibnamefont
  {Carlson}}, \bibinfo {author} {\bibfnamefont {S.}~\bibnamefont {Gandolfi}},
  \bibinfo {author} {\bibfnamefont {F.}~\bibnamefont {Pederiva}}, \bibinfo
  {author} {\bibfnamefont {S.~C.}\ \bibnamefont {Pieper}}, \bibinfo {author}
  {\bibfnamefont {R.}~\bibnamefont {Schiavilla}}, \bibinfo {author}
  {\bibfnamefont {K.~E.}\ \bibnamefont {Schmidt}},\ and\ \bibinfo {author}
  {\bibfnamefont {R.~B.}\ \bibnamefont {Wiringa}},\ }\bibfield  {title}
  {\bibinfo {title} {{Quantum Monte Carlo methods for nuclear physics}},\
  }\href {https://doi.org/10.1103/RevModPhys.87.1067} {\bibfield  {journal}
  {\bibinfo  {journal} {Rev. Mod. Phys.}\ }\textbf {\bibinfo {volume} {87}},\
  \bibinfo {pages} {1067} (\bibinfo {year} {2015})}\BibitemShut {NoStop}%
\bibitem [{\citenamefont {{Or{\'u}s}}(2019)}]{Orus:2019}%
  \BibitemOpen
  \bibfield  {author} {\bibinfo {author} {\bibfnamefont {R.}~\bibnamefont
  {{Or{\'u}s}}},\ }\bibfield  {title} {\bibinfo {title} {{Tensor networks for
  complex quantum systems}},\ }\href
  {https://doi.org/10.1038/s42254-019-0086-7} {\bibfield  {journal} {\bibinfo
  {journal} {Nature Reviews Physics}\ }\textbf {\bibinfo {volume} {1}},\
  \bibinfo {pages} {538} (\bibinfo {year} {2019})},\ \Eprint
  {https://arxiv.org/abs/1812.04011} {arXiv:1812.04011 [cond-mat.str-el]}
  \BibitemShut {NoStop}%
\bibitem [{\citenamefont {Bartlett}(1981)}]{Bartlett:1981}%
  \BibitemOpen
  \bibfield  {author} {\bibinfo {author} {\bibfnamefont {R.~J.}\ \bibnamefont
  {Bartlett}},\ }\bibfield  {title} {\bibinfo {title} {Many-body perturbation
  theory and coupled cluster theory for electron correlation in molecules},\
  }\href {https://doi.org/10.1146/annurev.pc.32.100181.002043} {\bibfield
  {journal} {\bibinfo  {journal} {Annual Review of Physical Chemistry}\
  }\textbf {\bibinfo {volume} {32}},\ \bibinfo {pages} {359} (\bibinfo {year}
  {1981})}\BibitemShut {NoStop}%
\bibitem [{\citenamefont {{Hagen}}\ \emph {et~al.}(2014)\citenamefont
  {{Hagen}}, \citenamefont {{Papenbrock}}, \citenamefont {{Hjorth-Jensen}},\
  and\ \citenamefont {{Dean}}}]{Hagen2014}%
  \BibitemOpen
  \bibfield  {author} {\bibinfo {author} {\bibfnamefont {G.}~\bibnamefont
  {{Hagen}}}, \bibinfo {author} {\bibfnamefont {T.}~\bibnamefont
  {{Papenbrock}}}, \bibinfo {author} {\bibfnamefont {M.}~\bibnamefont
  {{Hjorth-Jensen}}},\ and\ \bibinfo {author} {\bibfnamefont {D.~J.}\
  \bibnamefont {{Dean}}},\ }\bibfield  {title} {\bibinfo {title}
  {{Coupled-cluster computations of atomic nuclei}},\ }\href
  {https://doi.org/10.1088/0034-4885/77/9/096302} {\bibfield  {journal}
  {\bibinfo  {journal} {Reports on Progress in Physics}\ }\textbf {\bibinfo
  {volume} {77}},\ \bibinfo {eid} {096302} (\bibinfo {year} {2014})},\ \Eprint
  {https://arxiv.org/abs/1312.7872} {arXiv:1312.7872 [nucl-th]} \BibitemShut
  {NoStop}%
\bibitem [{\citenamefont {Ge}\ \emph {et~al.}(2019)\citenamefont {Ge},
  \citenamefont {Tura},\ and\ \citenamefont {Cirac}}]{Ge2018}%
  \BibitemOpen
  \bibfield  {author} {\bibinfo {author} {\bibfnamefont {Y.}~\bibnamefont
  {Ge}}, \bibinfo {author} {\bibfnamefont {J.}~\bibnamefont {Tura}},\ and\
  \bibinfo {author} {\bibfnamefont {J.~I.}\ \bibnamefont {Cirac}},\ }\bibfield
  {title} {\bibinfo {title} {Faster ground state preparation and high-precision
  ground energy estimation with fewer qubits},\ }\href
  {https://doi.org/https://doi.org/10.1063/1.5027484} {\bibfield  {journal}
  {\bibinfo  {journal} {J. Math. Phys.}\ }\textbf {\bibinfo {volume} {60}},\
  \bibinfo {pages} {022202} (\bibinfo {year} {2019})}\BibitemShut {NoStop}%
\bibitem [{\citenamefont {Berry}\ \emph {et~al.}(2007)\citenamefont {Berry},
  \citenamefont {Ahokas}, \citenamefont {Cleve},\ and\ \citenamefont
  {Sanders}}]{Berry_2006}%
  \BibitemOpen
  \bibfield  {author} {\bibinfo {author} {\bibfnamefont {D.~W.}\ \bibnamefont
  {Berry}}, \bibinfo {author} {\bibfnamefont {G.}~\bibnamefont {Ahokas}},
  \bibinfo {author} {\bibfnamefont {R.}~\bibnamefont {Cleve}},\ and\ \bibinfo
  {author} {\bibfnamefont {B.~C.}\ \bibnamefont {Sanders}},\ }\bibfield
  {title} {\bibinfo {title} {Efficient quantum algorithms for simulating sparse
  hamiltonians},\ }\href {https://doi.org/10.1007/s00220-006-0150-x} {\bibfield
   {journal} {\bibinfo  {journal} {Communications in Mathematical Physics}\
  }\textbf {\bibinfo {volume} {270}},\ \bibinfo {pages} {359} (\bibinfo {year}
  {2007})}\BibitemShut {NoStop}%
\bibitem [{\citenamefont {{Kandala}}\ \emph {et~al.}(2017)\citenamefont
  {{Kandala}}, \citenamefont {{Mezzacapo}}, \citenamefont {{Temme}},
  \citenamefont {{Takita}}, \citenamefont {{Brink}}, \citenamefont {{Chow}},\
  and\ \citenamefont {{Gambetta}}}]{Kandala:2017HE}%
  \BibitemOpen
  \bibfield  {author} {\bibinfo {author} {\bibfnamefont {A.}~\bibnamefont
  {{Kandala}}}, \bibinfo {author} {\bibfnamefont {A.}~\bibnamefont
  {{Mezzacapo}}}, \bibinfo {author} {\bibfnamefont {K.}~\bibnamefont
  {{Temme}}}, \bibinfo {author} {\bibfnamefont {M.}~\bibnamefont {{Takita}}},
  \bibinfo {author} {\bibfnamefont {M.}~\bibnamefont {{Brink}}}, \bibinfo
  {author} {\bibfnamefont {J.~M.}\ \bibnamefont {{Chow}}},\ and\ \bibinfo
  {author} {\bibfnamefont {J.~M.}\ \bibnamefont {{Gambetta}}},\ }\bibfield
  {title} {\bibinfo {title} {{Hardware-efficient variational quantum
  eigensolver for small molecules and quantum magnets}},\ }\href
  {https://doi.org/10.1038/nature23879} {\bibfield  {journal} {\bibinfo
  {journal} {\nat}\ }\textbf {\bibinfo {volume} {549}},\ \bibinfo {pages} {242}
  (\bibinfo {year} {2017})},\ \Eprint {https://arxiv.org/abs/1704.05018}
  {arXiv:1704.05018 [quant-ph]} \BibitemShut {NoStop}%
\bibitem [{\citenamefont {Tilly}\ \emph {et~al.}(2022)\citenamefont {Tilly}
  \emph {et~al.}}]{Tilly:2021}%
  \BibitemOpen
  \bibfield  {author} {\bibinfo {author} {\bibfnamefont {J.}~\bibnamefont
  {Tilly}} \emph {et~al.},\ }\bibfield  {title} {\bibinfo {title} {{The
  Variational Quantum Eigensolver: A review of methods and best practices}},\
  }\href {https://doi.org/10.1016/j.physrep.2022.08.003} {\bibfield  {journal}
  {\bibinfo  {journal} {Phys. Rept.}\ }\textbf {\bibinfo {volume} {986}},\
  \bibinfo {pages} {1} (\bibinfo {year} {2022})},\ \Eprint
  {https://arxiv.org/abs/2111.05176} {arXiv:2111.05176 [quant-ph]} \BibitemShut
  {NoStop}%
\bibitem [{\citenamefont {{Farhi}}\ \emph {et~al.}(2000)\citenamefont
  {{Farhi}}, \citenamefont {{Goldstone}}, \citenamefont {{Gutmann}},\ and\
  \citenamefont {{Sipser}}}]{Farhi2000}%
  \BibitemOpen
  \bibfield  {author} {\bibinfo {author} {\bibfnamefont {E.}~\bibnamefont
  {{Farhi}}}, \bibinfo {author} {\bibfnamefont {J.}~\bibnamefont
  {{Goldstone}}}, \bibinfo {author} {\bibfnamefont {S.}~\bibnamefont
  {{Gutmann}}},\ and\ \bibinfo {author} {\bibfnamefont {M.}~\bibnamefont
  {{Sipser}}},\ }\bibfield  {title} {\bibinfo {title} {{Quantum Computation by
  Adiabatic Evolution}},\ }\href@noop {} {\bibfield  {journal} {\bibinfo
  {journal} {arXiv e-prints}\ ,\ \bibinfo {eid} {quant-ph/0001106}} (\bibinfo
  {year} {2000})},\ \Eprint {https://arxiv.org/abs/quant-ph/0001106}
  {arXiv:quant-ph/0001106 [quant-ph]} \BibitemShut {NoStop}%
\bibitem [{\citenamefont {Albash}\ and\ \citenamefont
  {Lidar}(2018)}]{Albash2018}%
  \BibitemOpen
  \bibfield  {author} {\bibinfo {author} {\bibfnamefont {T.}~\bibnamefont
  {Albash}}\ and\ \bibinfo {author} {\bibfnamefont {D.~A.}\ \bibnamefont
  {Lidar}},\ }\bibfield  {title} {\bibinfo {title} {Adiabatic quantum
  computation},\ }\href {https://doi.org/10.1103/RevModPhys.90.015002}
  {\bibfield  {journal} {\bibinfo  {journal} {Rev. Mod. Phys.}\ }\textbf
  {\bibinfo {volume} {90}},\ \bibinfo {pages} {015002} (\bibinfo {year}
  {2018})}\BibitemShut {NoStop}%
\bibitem [{\citenamefont {Stannigel}\ \emph {et~al.}(2014)\citenamefont
  {Stannigel}, \citenamefont {Hauke}, \citenamefont {Marcos}, \citenamefont
  {Hafezi}, \citenamefont {Diehl}, \citenamefont {Dalmonte},\ and\
  \citenamefont {Zoller}}]{PhysRevLett.112.120406}%
  \BibitemOpen
  \bibfield  {author} {\bibinfo {author} {\bibfnamefont {K.}~\bibnamefont
  {Stannigel}}, \bibinfo {author} {\bibfnamefont {P.}~\bibnamefont {Hauke}},
  \bibinfo {author} {\bibfnamefont {D.}~\bibnamefont {Marcos}}, \bibinfo
  {author} {\bibfnamefont {M.}~\bibnamefont {Hafezi}}, \bibinfo {author}
  {\bibfnamefont {S.}~\bibnamefont {Diehl}}, \bibinfo {author} {\bibfnamefont
  {M.}~\bibnamefont {Dalmonte}},\ and\ \bibinfo {author} {\bibfnamefont
  {P.}~\bibnamefont {Zoller}},\ }\bibfield  {title} {\bibinfo {title}
  {Constrained dynamics via the zeno effect in quantum simulation: Implementing
  non-abelian lattice gauge theories with cold atoms},\ }\href
  {https://doi.org/10.1103/PhysRevLett.112.120406} {\bibfield  {journal}
  {\bibinfo  {journal} {Phys. Rev. Lett.}\ }\textbf {\bibinfo {volume} {112}},\
  \bibinfo {pages} {120406} (\bibinfo {year} {2014})}\BibitemShut {NoStop}%
\bibitem [{\citenamefont {Tran}\ \emph {et~al.}(2021)\citenamefont {Tran},
  \citenamefont {Su}, \citenamefont {Carney},\ and\ \citenamefont
  {Taylor}}]{PRXQuantum.2.010323}%
  \BibitemOpen
  \bibfield  {author} {\bibinfo {author} {\bibfnamefont {M.~C.}\ \bibnamefont
  {Tran}}, \bibinfo {author} {\bibfnamefont {Y.}~\bibnamefont {Su}}, \bibinfo
  {author} {\bibfnamefont {D.}~\bibnamefont {Carney}},\ and\ \bibinfo {author}
  {\bibfnamefont {J.~M.}\ \bibnamefont {Taylor}},\ }\bibfield  {title}
  {\bibinfo {title} {Faster digital quantum simulation by symmetry
  protection},\ }\href {https://doi.org/10.1103/PRXQuantum.2.010323} {\bibfield
   {journal} {\bibinfo  {journal} {PRX Quantum}\ }\textbf {\bibinfo {volume}
  {2}},\ \bibinfo {pages} {010323} (\bibinfo {year} {2021})}\BibitemShut
  {NoStop}%
\bibitem [{\citenamefont {Halimeh}\ \emph {et~al.}(2021)\citenamefont
  {Halimeh}, \citenamefont {Lang}, \citenamefont {Mildenberger}, \citenamefont
  {Jiang},\ and\ \citenamefont {Hauke}}]{PRXQuantum.2.040311}%
  \BibitemOpen
  \bibfield  {author} {\bibinfo {author} {\bibfnamefont {J.~C.}\ \bibnamefont
  {Halimeh}}, \bibinfo {author} {\bibfnamefont {H.}~\bibnamefont {Lang}},
  \bibinfo {author} {\bibfnamefont {J.}~\bibnamefont {Mildenberger}}, \bibinfo
  {author} {\bibfnamefont {Z.}~\bibnamefont {Jiang}},\ and\ \bibinfo {author}
  {\bibfnamefont {P.}~\bibnamefont {Hauke}},\ }\bibfield  {title} {\bibinfo
  {title} {Gauge-symmetry protection using single-body terms},\ }\href
  {https://doi.org/10.1103/PRXQuantum.2.040311} {\bibfield  {journal} {\bibinfo
   {journal} {PRX Quantum}\ }\textbf {\bibinfo {volume} {2}},\ \bibinfo {pages}
  {040311} (\bibinfo {year} {2021})}\BibitemShut {NoStop}%
\bibitem [{\citenamefont {Nguyen}\ \emph {et~al.}(2022)\citenamefont {Nguyen},
  \citenamefont {Tran}, \citenamefont {Zhu}, \citenamefont {Green},
  \citenamefont {Alderete}, \citenamefont {Davoudi},\ and\ \citenamefont
  {Linke}}]{PRXQuantum.3.020324}%
  \BibitemOpen
  \bibfield  {author} {\bibinfo {author} {\bibfnamefont {N.~H.}\ \bibnamefont
  {Nguyen}}, \bibinfo {author} {\bibfnamefont {M.~C.}\ \bibnamefont {Tran}},
  \bibinfo {author} {\bibfnamefont {Y.}~\bibnamefont {Zhu}}, \bibinfo {author}
  {\bibfnamefont {A.~M.}\ \bibnamefont {Green}}, \bibinfo {author}
  {\bibfnamefont {C.~H.}\ \bibnamefont {Alderete}}, \bibinfo {author}
  {\bibfnamefont {Z.}~\bibnamefont {Davoudi}},\ and\ \bibinfo {author}
  {\bibfnamefont {N.~M.}\ \bibnamefont {Linke}},\ }\bibfield  {title} {\bibinfo
  {title} {Digital quantum simulation of the {S}chwinger model and symmetry
  protection with trapped ions},\ }\href
  {https://doi.org/10.1103/PRXQuantum.3.020324} {\bibfield  {journal} {\bibinfo
   {journal} {PRX Quantum}\ }\textbf {\bibinfo {volume} {3}},\ \bibinfo {pages}
  {020324} (\bibinfo {year} {2022})}\BibitemShut {NoStop}%
\bibitem [{\citenamefont {Halimeh}\ \emph {et~al.}(2022)\citenamefont
  {Halimeh}, \citenamefont {Lang},\ and\ \citenamefont {Hauke}}]{Halimeh_2022}%
  \BibitemOpen
  \bibfield  {author} {\bibinfo {author} {\bibfnamefont {J.~C.}\ \bibnamefont
  {Halimeh}}, \bibinfo {author} {\bibfnamefont {H.}~\bibnamefont {Lang}},\ and\
  \bibinfo {author} {\bibfnamefont {P.}~\bibnamefont {Hauke}},\ }\bibfield
  {title} {\bibinfo {title} {Gauge protection in non-abelian lattice gauge
  theories},\ }\href {https://doi.org/10.1088/1367-2630/ac5564} {\bibfield
  {journal} {\bibinfo  {journal} {New Journal of Physics}\ }\textbf {\bibinfo
  {volume} {24}},\ \bibinfo {pages} {033015} (\bibinfo {year}
  {2022})}\BibitemShut {NoStop}%
\bibitem [{\citenamefont {Lacroix}(2020)}]{PhysRevLett.125.230502}%
  \BibitemOpen
  \bibfield  {author} {\bibinfo {author} {\bibfnamefont {D.}~\bibnamefont
  {Lacroix}},\ }\bibfield  {title} {\bibinfo {title} {Symmetry-assisted
  preparation of entangled many-body states on a quantum computer},\ }\href
  {https://doi.org/10.1103/PhysRevLett.125.230502} {\bibfield  {journal}
  {\bibinfo  {journal} {Phys. Rev. Lett.}\ }\textbf {\bibinfo {volume} {125}},\
  \bibinfo {pages} {230502} (\bibinfo {year} {2020})}\BibitemShut {NoStop}%
\bibitem [{\citenamefont {Siwach}\ and\ \citenamefont
  {Lacroix}(2021)}]{PhysRevA.104.062435}%
  \BibitemOpen
  \bibfield  {author} {\bibinfo {author} {\bibfnamefont {P.}~\bibnamefont
  {Siwach}}\ and\ \bibinfo {author} {\bibfnamefont {D.}~\bibnamefont
  {Lacroix}},\ }\bibfield  {title} {\bibinfo {title} {Filtering states with
  total spin on a quantum computer},\ }\href
  {https://doi.org/10.1103/PhysRevA.104.062435} {\bibfield  {journal} {\bibinfo
   {journal} {Phys. Rev. A}\ }\textbf {\bibinfo {volume} {104}},\ \bibinfo
  {pages} {062435} (\bibinfo {year} {2021})}\BibitemShut {NoStop}%
\bibitem [{\citenamefont {Ruiz~Guzman}\ and\ \citenamefont
  {Lacroix}(2022)}]{PhysRevC.105.024324}%
  \BibitemOpen
  \bibfield  {author} {\bibinfo {author} {\bibfnamefont {E.~A.}\ \bibnamefont
  {Ruiz~Guzman}}\ and\ \bibinfo {author} {\bibfnamefont {D.}~\bibnamefont
  {Lacroix}},\ }\bibfield  {title} {\bibinfo {title} {Accessing ground-state
  and excited-state energies in a many-body system after symmetry restoration
  using quantum computers},\ }\href
  {https://doi.org/10.1103/PhysRevC.105.024324} {\bibfield  {journal} {\bibinfo
   {journal} {Phys. Rev. C}\ }\textbf {\bibinfo {volume} {105}},\ \bibinfo
  {pages} {024324} (\bibinfo {year} {2022})}\BibitemShut {NoStop}%
\bibitem [{\citenamefont {{Motta}}\ \emph {et~al.}(2020)\citenamefont
  {{Motta}}, \citenamefont {{Sun}}, \citenamefont {{Tan}}, \citenamefont
  {{O'Rourke}}, \citenamefont {{Ye}}, \citenamefont {{Minnich}}, \citenamefont
  {{Brand{\~a}o}},\ and\ \citenamefont {{Chan}}}]{Motta:2020}%
  \BibitemOpen
  \bibfield  {author} {\bibinfo {author} {\bibfnamefont {M.}~\bibnamefont
  {{Motta}}}, \bibinfo {author} {\bibfnamefont {C.}~\bibnamefont {{Sun}}},
  \bibinfo {author} {\bibfnamefont {A.~T.~K.}\ \bibnamefont {{Tan}}}, \bibinfo
  {author} {\bibfnamefont {M.~J.}\ \bibnamefont {{O'Rourke}}}, \bibinfo
  {author} {\bibfnamefont {E.}~\bibnamefont {{Ye}}}, \bibinfo {author}
  {\bibfnamefont {A.~J.}\ \bibnamefont {{Minnich}}}, \bibinfo {author}
  {\bibfnamefont {F.~G.~S.~L.}\ \bibnamefont {{Brand{\~a}o}}},\ and\ \bibinfo
  {author} {\bibfnamefont {G.~K.-L.}\ \bibnamefont {{Chan}}},\ }\bibfield
  {title} {\bibinfo {title} {{Determining eigenstates and thermal states on a
  quantum computer using quantum imaginary time evolution}},\ }\href
  {https://doi.org/10.1038/s41567-019-0704-4} {\bibfield  {journal} {\bibinfo
  {journal} {Nature Physics}\ }\textbf {\bibinfo {volume} {16}},\ \bibinfo
  {pages} {205} (\bibinfo {year} {2020})},\ \Eprint
  {https://arxiv.org/abs/1901.07653} {arXiv:1901.07653 [quant-ph]} \BibitemShut
  {NoStop}%
\bibitem [{\citenamefont {Kosugi}\ \emph {et~al.}(2022)\citenamefont {Kosugi},
  \citenamefont {Nishiya}, \citenamefont {Nishi},\ and\ \citenamefont
  {Matsushita}}]{PhysRevResearch.4.033121}%
  \BibitemOpen
  \bibfield  {author} {\bibinfo {author} {\bibfnamefont {T.}~\bibnamefont
  {Kosugi}}, \bibinfo {author} {\bibfnamefont {Y.}~\bibnamefont {Nishiya}},
  \bibinfo {author} {\bibfnamefont {H.}~\bibnamefont {Nishi}},\ and\ \bibinfo
  {author} {\bibfnamefont {Y.-i.}\ \bibnamefont {Matsushita}},\ }\bibfield
  {title} {\bibinfo {title} {Imaginary-time evolution using forward and
  backward real-time evolution with a single ancilla: First-quantized
  eigensolver algorithm for quantum chemistry},\ }\href
  {https://doi.org/10.1103/PhysRevResearch.4.033121} {\bibfield  {journal}
  {\bibinfo  {journal} {Phys. Rev. Research}\ }\textbf {\bibinfo {volume}
  {4}},\ \bibinfo {pages} {033121} (\bibinfo {year} {2022})}\BibitemShut
  {NoStop}%
\bibitem [{\citenamefont {Turro}\ \emph {et~al.}(2022)\citenamefont {Turro},
  \citenamefont {Roggero}, \citenamefont {Amitrano}, \citenamefont {Luchi},
  \citenamefont {Wendt}, \citenamefont {Dubois}, \citenamefont {Quaglioni},\
  and\ \citenamefont {Pederiva}}]{Turro2022}%
  \BibitemOpen
  \bibfield  {author} {\bibinfo {author} {\bibfnamefont {F.}~\bibnamefont
  {Turro}}, \bibinfo {author} {\bibfnamefont {A.}~\bibnamefont {Roggero}},
  \bibinfo {author} {\bibfnamefont {V.}~\bibnamefont {Amitrano}}, \bibinfo
  {author} {\bibfnamefont {P.}~\bibnamefont {Luchi}}, \bibinfo {author}
  {\bibfnamefont {K.~A.}\ \bibnamefont {Wendt}}, \bibinfo {author}
  {\bibfnamefont {J.~L.}\ \bibnamefont {Dubois}}, \bibinfo {author}
  {\bibfnamefont {S.}~\bibnamefont {Quaglioni}},\ and\ \bibinfo {author}
  {\bibfnamefont {F.}~\bibnamefont {Pederiva}},\ }\bibfield  {title} {\bibinfo
  {title} {{Imaginary-time propagation on a quantum chip}},\ }\href
  {https://doi.org/10.1103/PhysRevA.105.022440} {\bibfield  {journal} {\bibinfo
   {journal} {Physical Review A}\ }\textbf {\bibinfo {volume} {105}},\ \bibinfo
  {pages} {022440} (\bibinfo {year} {2022})}\BibitemShut {NoStop}%
\bibitem [{\citenamefont {Jouzdani}\ \emph {et~al.}(2022)\citenamefont
  {Jouzdani}, \citenamefont {Johnson}, \citenamefont {Mucciolo},\ and\
  \citenamefont {Stetcu}}]{Jouzdani2022}%
  \BibitemOpen
  \bibfield  {author} {\bibinfo {author} {\bibfnamefont {P.}~\bibnamefont
  {Jouzdani}}, \bibinfo {author} {\bibfnamefont {C.~W.}\ \bibnamefont
  {Johnson}}, \bibinfo {author} {\bibfnamefont {E.~R.}\ \bibnamefont
  {Mucciolo}},\ and\ \bibinfo {author} {\bibfnamefont {I.}~\bibnamefont
  {Stetcu}},\ }\bibfield  {title} {\bibinfo {title} {Alternative approach to
  quantum imaginary time evolution},\ }\href
  {https://doi.org/10.1103/PhysRevA.106.062435} {\bibfield  {journal} {\bibinfo
   {journal} {Phys. Rev. A}\ }\textbf {\bibinfo {volume} {106}},\ \bibinfo
  {pages} {062435} (\bibinfo {year} {2022})}\BibitemShut {NoStop}%
\bibitem [{\citenamefont {Holmes}\ \emph {et~al.}(2022)\citenamefont {Holmes},
  \citenamefont {Muraleedharan}, \citenamefont {Somma}, \citenamefont
  {Subasi},\ and\ \citenamefont
  {{\c{S}}ahino{\u{g}}lu}}]{Holmes2022quantumalgorithms}%
  \BibitemOpen
  \bibfield  {author} {\bibinfo {author} {\bibfnamefont {Z.}~\bibnamefont
  {Holmes}}, \bibinfo {author} {\bibfnamefont {G.}~\bibnamefont
  {Muraleedharan}}, \bibinfo {author} {\bibfnamefont {R.~D.}\ \bibnamefont
  {Somma}}, \bibinfo {author} {\bibfnamefont {Y.}~\bibnamefont {Subasi}},\ and\
  \bibinfo {author} {\bibfnamefont {B.}~\bibnamefont {{\c{S}}ahino{\u{g}}lu}},\
  }\bibfield  {title} {\bibinfo {title} {Quantum algorithms from fluctuation
  theorems: {T}hermal-state preparation},\ }\href
  {https://doi.org/10.22331/q-2022-10-06-825} {\bibfield  {journal} {\bibinfo
  {journal} {{Quantum}}\ }\textbf {\bibinfo {volume} {6}},\ \bibinfo {pages}
  {825} (\bibinfo {year} {2022})}\BibitemShut {NoStop}%
\bibitem [{\citenamefont {Dong}\ \emph {et~al.}(2022)\citenamefont {Dong},
  \citenamefont {Lin},\ and\ \citenamefont {Tong}}]{Dong2022}%
  \BibitemOpen
  \bibfield  {author} {\bibinfo {author} {\bibfnamefont {Y.}~\bibnamefont
  {Dong}}, \bibinfo {author} {\bibfnamefont {L.}~\bibnamefont {Lin}},\ and\
  \bibinfo {author} {\bibfnamefont {Y.}~\bibnamefont {Tong}},\ }\bibfield
  {title} {\bibinfo {title} {Ground-state preparation and energy estimation on
  early fault-tolerant quantum computers via quantum eigenvalue transformation
  of unitary matrices},\ }\href {https://doi.org/10.1103/PRXQuantum.3.040305}
  {\bibfield  {journal} {\bibinfo  {journal} {PRX Quantum}\ }\textbf {\bibinfo
  {volume} {3}},\ \bibinfo {pages} {040305} (\bibinfo {year}
  {2022})}\BibitemShut {NoStop}%
\bibitem [{\citenamefont {Keen}\ \emph {et~al.}(2021)\citenamefont {Keen},
  \citenamefont {Dumitrescu},\ and\ \citenamefont {Wang}}]{Keen2021}%
  \BibitemOpen
  \bibfield  {author} {\bibinfo {author} {\bibfnamefont {T.}~\bibnamefont
  {Keen}}, \bibinfo {author} {\bibfnamefont {E.}~\bibnamefont {Dumitrescu}},\
  and\ \bibinfo {author} {\bibfnamefont {Y.}~\bibnamefont {Wang}},\ }\bibfield
  {title} {\bibinfo {title} {{Quantum Algorithms for Ground-State Preparation
  and Green's Function Calculation}},\ }\href@noop {} {\bibfield  {journal}
  {\bibinfo  {journal} {arXiv e-prints}\ ,\ \bibinfo {eid} {arXiv:2112.05731}}
  (\bibinfo {year} {2021})},\ \Eprint {https://arxiv.org/abs/2112.05731}
  {arXiv:2112.05731 [quant-ph]} \BibitemShut {NoStop}%
\bibitem [{\citenamefont {Choi}\ \emph {et~al.}(2021)\citenamefont {Choi},
  \citenamefont {Lee}, \citenamefont {Bonitati}, \citenamefont {Qian},\ and\
  \citenamefont {Watkins}}]{Choi2021}%
  \BibitemOpen
  \bibfield  {author} {\bibinfo {author} {\bibfnamefont {K.}~\bibnamefont
  {Choi}}, \bibinfo {author} {\bibfnamefont {D.}~\bibnamefont {Lee}}, \bibinfo
  {author} {\bibfnamefont {J.}~\bibnamefont {Bonitati}}, \bibinfo {author}
  {\bibfnamefont {Z.}~\bibnamefont {Qian}},\ and\ \bibinfo {author}
  {\bibfnamefont {J.}~\bibnamefont {Watkins}},\ }\bibfield  {title} {\bibinfo
  {title} {Rodeo algorithm for quantum computing},\ }\href
  {https://doi.org/10.1103/PhysRevLett.127.040505} {\bibfield  {journal}
  {\bibinfo  {journal} {Phys. Rev. Lett.}\ }\textbf {\bibinfo {volume} {127}},\
  \bibinfo {pages} {040505} (\bibinfo {year} {2021})}\BibitemShut {NoStop}%
\bibitem [{\citenamefont {Roggero}\ and\ \citenamefont
  {Carlson}(2019)}]{Rogerro2019}%
  \BibitemOpen
  \bibfield  {author} {\bibinfo {author} {\bibfnamefont {A.}~\bibnamefont
  {Roggero}}\ and\ \bibinfo {author} {\bibfnamefont {J.}~\bibnamefont
  {Carlson}},\ }\bibfield  {title} {\bibinfo {title} {Dynamic linear response
  quantum algorithm},\ }\href {https://doi.org/10.1103/PhysRevC.100.034610}
  {\bibfield  {journal} {\bibinfo  {journal} {Phys. Rev. C}\ }\textbf {\bibinfo
  {volume} {100}},\ \bibinfo {pages} {034610} (\bibinfo {year}
  {2019})}\BibitemShut {NoStop}%
\bibitem [{sup(2022)}]{supplement}%
  \BibitemOpen
  \href@noop {} {\bibinfo {title} {{See Supplemental Material for more detailed
  explanations and examples not shown in the text.}}} (\bibinfo {year}
  {2022})\BibitemShut {NoStop}%
\bibitem [{\citenamefont {Somma}\ \emph {et~al.}(2002)\citenamefont {Somma},
  \citenamefont {Ortiz}, \citenamefont {Gubernatis}, \citenamefont {Knill},\
  and\ \citenamefont {Laflamme}}]{Soma-SpectrumState}%
  \BibitemOpen
  \bibfield  {author} {\bibinfo {author} {\bibfnamefont {R.}~\bibnamefont
  {Somma}}, \bibinfo {author} {\bibfnamefont {G.}~\bibnamefont {Ortiz}},
  \bibinfo {author} {\bibfnamefont {J.~E.}\ \bibnamefont {Gubernatis}},
  \bibinfo {author} {\bibfnamefont {E.}~\bibnamefont {Knill}},\ and\ \bibinfo
  {author} {\bibfnamefont {R.}~\bibnamefont {Laflamme}},\ }\bibfield  {title}
  {\bibinfo {title} {Simulating physical phenomena by quantum networks},\
  }\href {https://doi.org/10.1103/PhysRevA.65.042323} {\bibfield  {journal}
  {\bibinfo  {journal} {Phys. Rev. A}\ }\textbf {\bibinfo {volume} {65}},\
  \bibinfo {pages} {042323} (\bibinfo {year} {2002})}\BibitemShut {NoStop}%
\bibitem [{\citenamefont {Kliuchnikov}\ \emph {et~al.}(2016)\citenamefont
  {Kliuchnikov}, \citenamefont {Maslov},\ and\ \citenamefont
  {Mosca}}]{Kliuchnikov2016}%
  \BibitemOpen
  \bibfield  {author} {\bibinfo {author} {\bibfnamefont {V.}~\bibnamefont
  {Kliuchnikov}}, \bibinfo {author} {\bibfnamefont {D.}~\bibnamefont
  {Maslov}},\ and\ \bibinfo {author} {\bibfnamefont {M.}~\bibnamefont
  {Mosca}},\ }\bibfield  {title} {\bibinfo {title} {Practical approximation of
  single-qubit unitaries by single-qubit quantum clifford and t circuits},\
  }\href {https://doi.org/10.1109/TC.2015.2409842} {\bibfield  {journal}
  {\bibinfo  {journal} {IEEE Transactions on Computers}\ }\textbf {\bibinfo
  {volume} {65}},\ \bibinfo {pages} {161} (\bibinfo {year} {2016})}\BibitemShut
  {NoStop}%
\bibitem [{\citenamefont {Gheorghiu}\ \emph {et~al.}(2022)\citenamefont
  {Gheorghiu}, \citenamefont {Mosca},\ and\ \citenamefont
  {Mukhopadhyay}}]{Gheorghiu2022}%
  \BibitemOpen
  \bibfield  {author} {\bibinfo {author} {\bibfnamefont {V.}~\bibnamefont
  {Gheorghiu}}, \bibinfo {author} {\bibfnamefont {M.}~\bibnamefont {Mosca}},\
  and\ \bibinfo {author} {\bibfnamefont {P.}~\bibnamefont {Mukhopadhyay}},\
  }\bibfield  {title} {\bibinfo {title} {T-count and t-depth of any multi-qubit
  unitary},\ }\bibfield  {journal} {\bibinfo  {journal} {npj Quantum
  Information}\ }\textbf {\bibinfo {volume} {141}},\ \href
  {https://doi.org/10.1038/s41534-022-00651-y} {10.1038/s41534-022-00651-y}
  (\bibinfo {year} {2022})\BibitemShut {NoStop}%
\bibitem [{\citenamefont {Stetcu}\ \emph {et~al.}(2022)\citenamefont {Stetcu},
  \citenamefont {Baroni},\ and\ \citenamefont {Carlson}}]{Stetcu2022var}%
  \BibitemOpen
  \bibfield  {author} {\bibinfo {author} {\bibfnamefont {I.}~\bibnamefont
  {Stetcu}}, \bibinfo {author} {\bibfnamefont {A.}~\bibnamefont {Baroni}},\
  and\ \bibinfo {author} {\bibfnamefont {J.}~\bibnamefont {Carlson}},\
  }\bibfield  {title} {\bibinfo {title} {Variational approaches to constructing
  the many-body nuclear ground state for quantum computing},\ }\href
  {https://doi.org/10.1103/PhysRevC.105.064308} {\bibfield  {journal} {\bibinfo
   {journal} {Phys. Rev. C}\ }\textbf {\bibinfo {volume} {105}},\ \bibinfo
  {pages} {064308} (\bibinfo {year} {2022})}\BibitemShut {NoStop}%
\bibitem [{\citenamefont {Kiss}\ \emph {et~al.}(2022)\citenamefont {Kiss},
  \citenamefont {Grossi}, \citenamefont {Lougovski}, \citenamefont {Sanchez},
  \citenamefont {Vallecorsa},\ and\ \citenamefont
  {Papenbrock}}]{PhysRevC.106.034325}%
  \BibitemOpen
  \bibfield  {author} {\bibinfo {author} {\bibfnamefont {O.}~\bibnamefont
  {Kiss}}, \bibinfo {author} {\bibfnamefont {M.}~\bibnamefont {Grossi}},
  \bibinfo {author} {\bibfnamefont {P.}~\bibnamefont {Lougovski}}, \bibinfo
  {author} {\bibfnamefont {F.}~\bibnamefont {Sanchez}}, \bibinfo {author}
  {\bibfnamefont {S.}~\bibnamefont {Vallecorsa}},\ and\ \bibinfo {author}
  {\bibfnamefont {T.}~\bibnamefont {Papenbrock}},\ }\bibfield  {title}
  {\bibinfo {title} {Quantum computing of the $^{6}\mathrm{Li}$ nucleus via
  ordered unitary coupled clusters},\ }\href
  {https://doi.org/10.1103/PhysRevC.106.034325} {\bibfield  {journal} {\bibinfo
   {journal} {Phys. Rev. C}\ }\textbf {\bibinfo {volume} {106}},\ \bibinfo
  {pages} {034325} (\bibinfo {year} {2022})}\BibitemShut {NoStop}%
\bibitem [{\citenamefont {Cohen}\ and\ \citenamefont
  {Kurath}(1965)}]{COHEN19651}%
  \BibitemOpen
  \bibfield  {author} {\bibinfo {author} {\bibfnamefont {S.}~\bibnamefont
  {Cohen}}\ and\ \bibinfo {author} {\bibfnamefont {D.}~\bibnamefont {Kurath}},\
  }\bibfield  {title} {\bibinfo {title} {Effective interactions for the 1p
  shell},\ }\href
  {https://doi.org/https://doi.org/10.1016/0029-5582(65)90148-3} {\bibfield
  {journal} {\bibinfo  {journal} {Nuclear Physics}\ }\textbf {\bibinfo {volume}
  {73}},\ \bibinfo {pages} {1} (\bibinfo {year} {1965})}\BibitemShut {NoStop}%
\end{thebibliography}%


\begin{thebibliography}{3}%
\makeatletter
\providecommand \@ifxundefined [1]{%
 \@ifx{#1\undefined}
}%
\providecommand \@ifnum [1]{%
 \ifnum #1\expandafter \@firstoftwo
 \else \expandafter \@secondoftwo
 \fi
}%
\providecommand \@ifx [1]{%
 \ifx #1\expandafter \@firstoftwo
 \else \expandafter \@secondoftwo
 \fi
}%
\providecommand \natexlab [1]{#1}%
\providecommand \enquote  [1]{``#1''}%
\providecommand \bibnamefont  [1]{#1}%
\providecommand \bibfnamefont [1]{#1}%
\providecommand \citenamefont [1]{#1}%
\providecommand \href@noop [0]{\@secondoftwo}%
\providecommand \href [0]{\begingroup \@sanitize@url \@href}%
\providecommand \@href[1]{\@@startlink{#1}\@@href}%
\providecommand \@@href[1]{\endgroup#1\@@endlink}%
\providecommand \@sanitize@url [0]{\catcode `\\12\catcode `\$12\catcode
  `\&12\catcode `\#12\catcode `\^12\catcode `\_12\catcode `\%12\relax}%
\providecommand \@@startlink[1]{}%
\providecommand \@@endlink[0]{}%
\providecommand \url  [0]{\begingroup\@sanitize@url \@url }%
\providecommand \@url [1]{\endgroup\@href {#1}{\urlprefix }}%
\providecommand \urlprefix  [0]{URL }%
\providecommand \Eprint [0]{\href }%
\providecommand \doibase [0]{https://doi.org/}%
\providecommand \selectlanguage [0]{\@gobble}%
\providecommand \bibinfo  [0]{\@secondoftwo}%
\providecommand \bibfield  [0]{\@secondoftwo}%
\providecommand \translation [1]{[#1]}%
\providecommand \BibitemOpen [0]{}%
\providecommand \bibitemStop [0]{}%
\providecommand \bibitemNoStop [0]{.\EOS\space}%
\providecommand \EOS [0]{\spacefactor3000\relax}%
\providecommand \BibitemShut  [1]{\csname bibitem#1\endcsname}%
\let\auto@bib@innerbib\@empty
\bibitem [{\citenamefont {Wildenthal}(1984)}]{WILDENTHAL19845}%
  \BibitemOpen
  \bibfield  {author} {\bibinfo {author} {\bibfnamefont {B.}~\bibnamefont
  {Wildenthal}},\ }\bibfield  {title} {\bibinfo {title} {Empirical strengths of
  spin operators in nuclei},\ }\href
  {https://doi.org/https://doi.org/10.1016/0146-6410(84)90011-5} {\bibfield
  {journal} {\bibinfo  {journal} {Progress in Particle and Nuclear Physics}\
  }\textbf {\bibinfo {volume} {11}},\ \bibinfo {pages} {5} (\bibinfo {year}
  {1984})}\BibitemShut {NoStop}%
\bibitem [{\citenamefont {Brown}\ and\ \citenamefont
  {Richter}(2006)}]{PhysRevC.74.034315}%
  \BibitemOpen
  \bibfield  {author} {\bibinfo {author} {\bibfnamefont {B.~A.}\ \bibnamefont
  {Brown}}\ and\ \bibinfo {author} {\bibfnamefont {W.~A.}\ \bibnamefont
  {Richter}},\ }\bibfield  {title} {\bibinfo {title} {{New ``USD'' Hamiltonians
  for the $\mathit{sd}$ shell}},\ }\href
  {https://doi.org/10.1103/PhysRevC.74.034315} {\bibfield  {journal} {\bibinfo
  {journal} {Phys. Rev. C}\ }\textbf {\bibinfo {volume} {74}},\ \bibinfo
  {pages} {034315} (\bibinfo {year} {2006})}\BibitemShut {NoStop}%
\bibitem [{\citenamefont {Cohen}\ and\ \citenamefont
  {Kurath}(1965)}]{COHEN19651}%
  \BibitemOpen
  \bibfield  {author} {\bibinfo {author} {\bibfnamefont {S.}~\bibnamefont
  {Cohen}}\ and\ \bibinfo {author} {\bibfnamefont {D.}~\bibnamefont {Kurath}},\
  }\bibfield  {title} {\bibinfo {title} {Effective interactions for the 1p
  shell},\ }\href
  {https://doi.org/https://doi.org/10.1016/0029-5582(65)90148-3} {\bibfield
  {journal} {\bibinfo  {journal} {Nuclear Physics}\ }\textbf {\bibinfo {volume}
  {73}},\ \bibinfo {pages} {1} (\bibinfo {year} {1965})}\BibitemShut {NoStop}%
\end{thebibliography}%

\end{document}


\onecolumngrid

    
\section{Projection Algorithm for State Preparation on Quantum Computers \\
Supplemental material}
\begin{center}
    I. Stetcu, A. Baroni, and J. Carlson
\end{center}

\subsection{Algorithmic Details and applications to the Heisenberg model}

In this supplemental material we illustrate
further details of the algorithm. We first present some details that are being exploited to perform the projection. We start again with Eq. (1) in the main paper
\begin{equation}
    |\psi(t)\rangle  =  \cos(t_i O+\delta_i)  |\psi\rangle \otimes |0\rangle_a 
    + \sin(t_i O+\delta_i)|\psi\rangle \otimes |1\rangle_a.
\label{eq:timeevolve-suppl}
\end{equation}
The operator $O$ and trial state $\psi$ can be both expanded using the spectrum $o_n$ and eigenstates $|v_n^o\rangle$ of the operator $O$, so that in this representation Eq. (\ref{eq:timeevolve-suppl}) becomes
\begin{eqnarray}
    |\psi(t) \rangle && =  \sum_{n}\cos(t_i o_n+\delta_i) |v_n^o\rangle \langle v_n^o |\psi\rangle \otimes |0\rangle_a \nonumber \\
    + && \sum_{n} \sin(t_i o_n+\delta_i)|v_n^o\rangle \langle v_n^o |\psi\rangle \otimes |1\rangle_a.
\end{eqnarray}
The eigenvectors $|v_n^o\rangle$ do not need to be known, but it helps if the eigenvalues $o_n$ are known, as in this case one can chose efficiently the times $t_i$ so that after each measurement of the ancilla qubit $a$ in state 0, we remove all the contributions from a set of eigenvalues. Note that if the target eigenvalue is not zero, we can always shift the spectrum $o_n$ by the desired value $o_\mathrm{target}$. For our algorithm, this reduces to evolving with the operator $O-o_\mathrm{target}$. Alternately, we can always use phases to get rid of states with unwanted symmetries. For example, if we target a spin $J=2$ state, and we want to remove $J=0$, we can chose $t_1=\pi/12$ and $\delta_1=-\pi/2$. However, this approach seems to be less efficient at removing the unwanted states than the simple shift.

\begin{figure}[hb]
\includegraphics[scale=0.5]{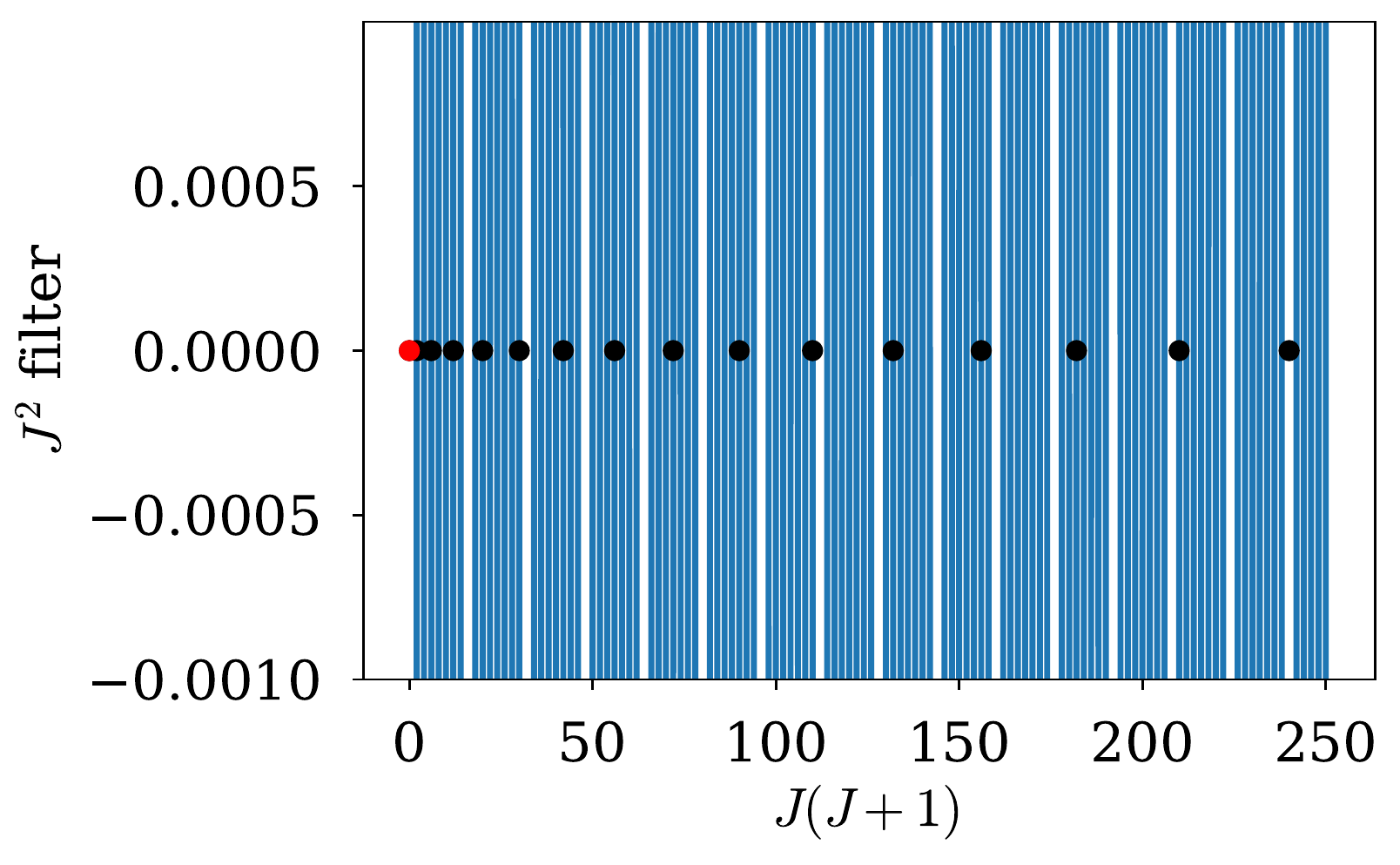}
\caption{Filter function after 3 iterations of 
Eq. (1) in the main text to project out $J^2=0$ state.
The black points in the figure are the values of
$J(J+1)$ we wish to filter
out, while the blue curve is the filter function. The red point represents the $J=0$ state.}
\label{fig:jsquaredfilter}
\end{figure}

Initially we describe in a bit more detail the
efficiency and impact of quantum number projection.
In Figure \ref{fig:jsquaredfilter} we show the
filter function after three iterations of the
Eq. (1) in the main text with times $\pi/4$, $\pi/8$ and $\pi/16$ to produce a filter that
eliminates all states with $J \leq 10$. In reality, the first $J$ that is not eliminated by this choice is $J=15$.

For the Heisenberg model above we considered an initial Neel state with up and down spins on the even and odd sublattices. We have also considered 
a very poor initial trial state that
has amplitude 1 in the ground state and amplitudes
randomly chosen between -1 and 1 for all other
states, with the whole state  then normalized to magnitude 1.
The initial state has a large $\langle J^2 \rangle$
and a large energy near the middle of the entire 
spectrum because of the random initial state.
The maximum possible $J$ in this system is $J=N/2 = 8$.
To project out all the $J>0$ states requires
three iterations, the last one would be sufficient to
project out higher $J$ on a bigger lattice.

\begin{figure}
\includegraphics[scale=0.4]{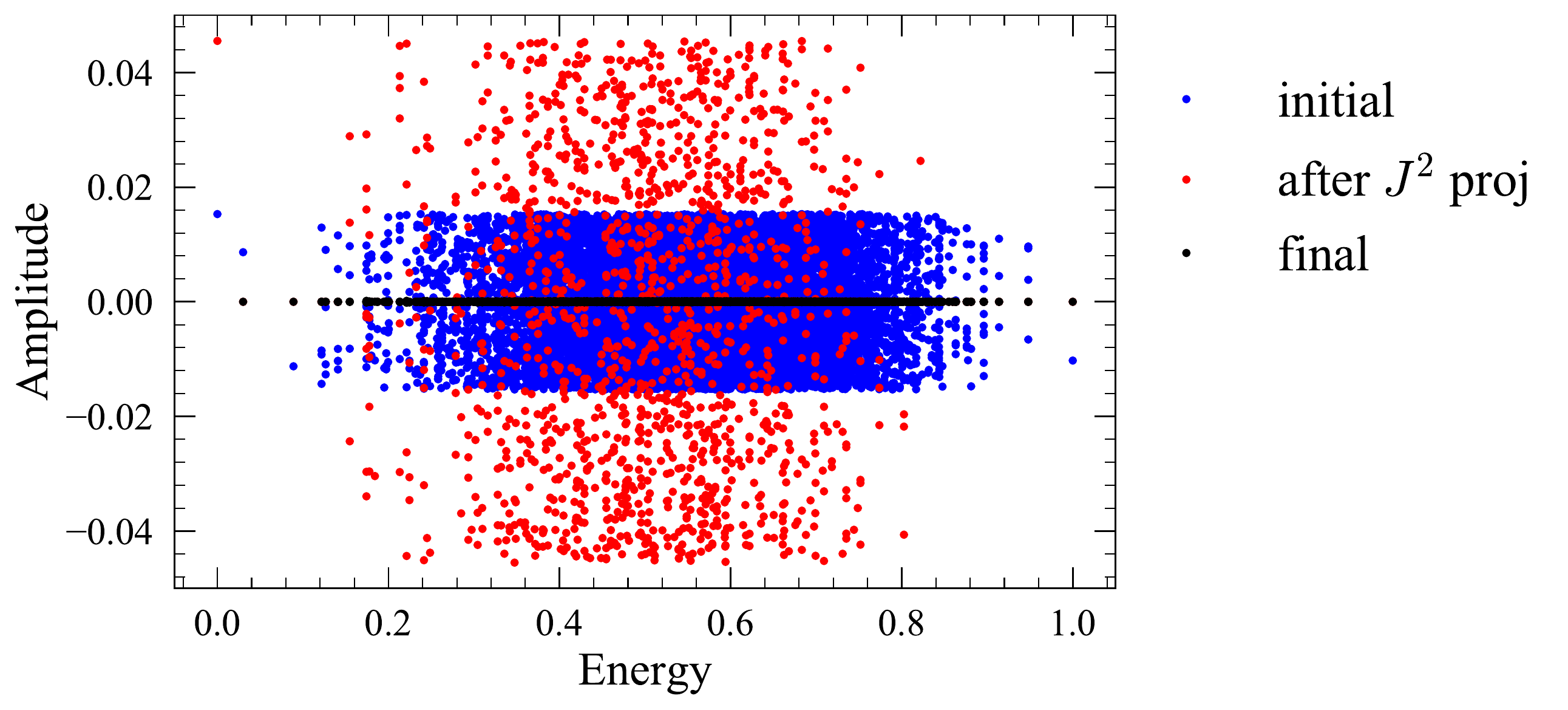}
\caption{Initial state amplitudes for the Heisenberg problem, those after
projection to $J=0$, and final amplitudes after energy projection. The final projected amplitude in
the ground state is near 1 and is not shown.}
\label{fig:heisenberg-amplitudes}
\end{figure}

In Fig. \ref{fig:heisenberg-amplitudes} we show
the amplitudes of the random initial state,
the state after $J^2$ projection and the state
after an additional energy projection.
The $J^2$ projection eliminates the large
number of states with $\langle J^2 \rangle > 0$,
and amplifies the remaining amplitude because
of the unitarity requirement.  

The energy projection after $J^2$ projection is
very effective here.  The original gap in this
problem is about 0.03, but increases to 0.15
with $J^2$ projection.  A similar number of measurements (7) and total time projection 
$(t = \pi/ \Delta )$ is required
to isolate the ground state from this random
initial state, though it is not as accurate
as starting from the Neel state. Exponential 
convergence with time is retained, however.

To optimize the times and phases, we consider
an initial state with all equal amplitudes,
with a finite fraction of amplitudes
at zero energy (in this case 50 out of 1000), and
the remainder distributed with energies from the
gap to one on a grid but randomly displaced within
a grid spacing to avoid artificial correlations.
We then define a target function that is one
at zero energy and zero at all the energies from the
gap to one.  We then calculate the impact of the
filter function [Eq. (1) in the main text] 
for a set of times and phases. We optimize
the magnitude of these times and phases to produce
the smallest mean squared difference between the
state produced by the algorithm acting on the initial
state and the target state all at energy zero.
The resulting filter function for an assumed gap
of 0.2 is shown in Fig. 1 in the main text.

\begin{figure}
\includegraphics[scale=0.4]{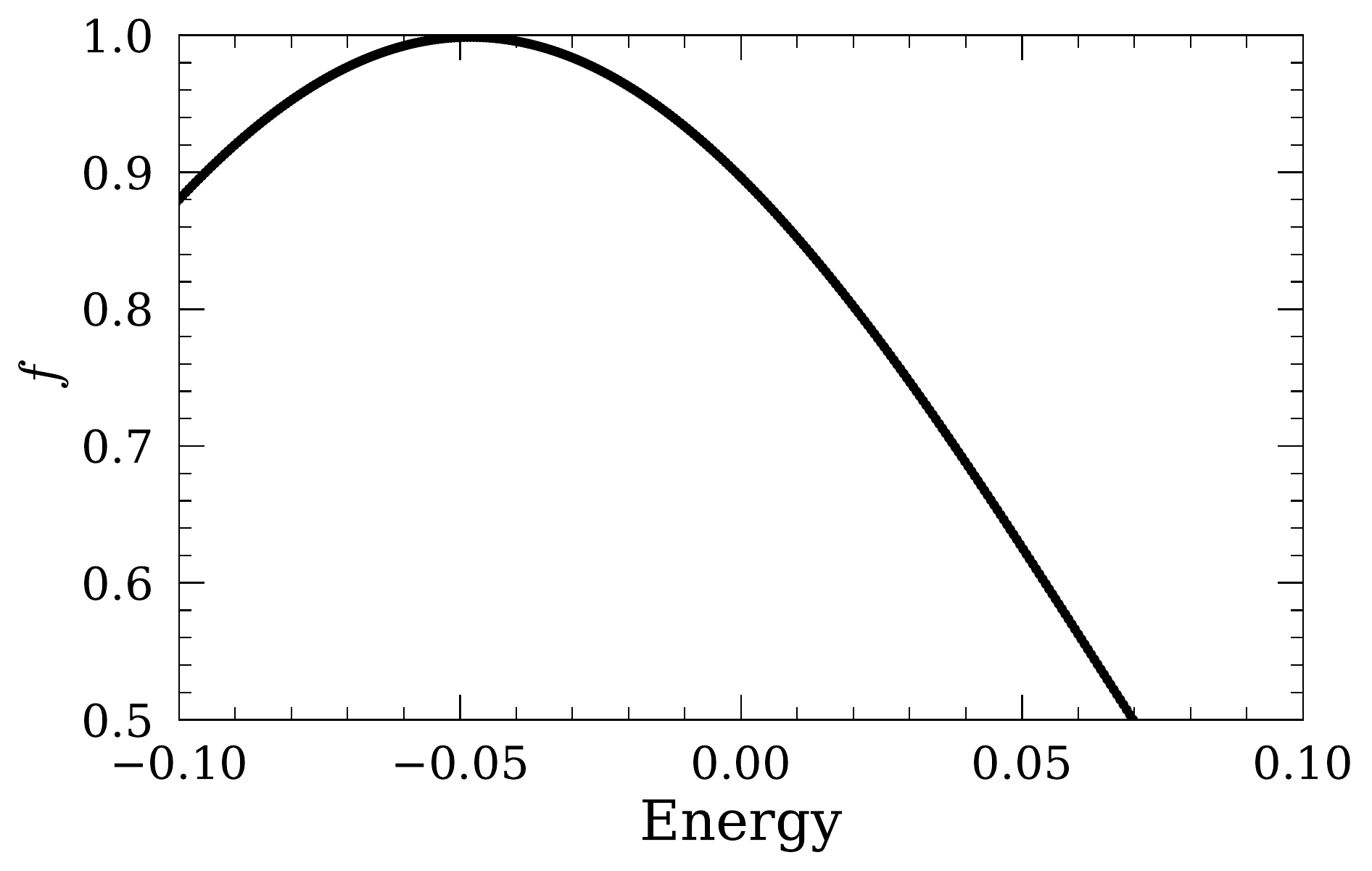}
\caption{Filter function in Fig. 1 of the main text 
zoomed in near zero energy.}
\label{fig:filternearzero}
\end{figure}

In figure \ref{fig:filternearzero} we show
the filter function near zero energy. Note because
of the additional phases the filter function does
not have zero slope near the origin. This reduces
the impact of uncertainties in the knowledge of the
ground state energy and the gap. In addition the phases
enable better convergence for systems with small gaps. All that is
required for convergence is that the filter function
at the true ground state energy is larger than the
filter function at the true first excited state.

\begin{figure}
\includegraphics[scale=0.4]{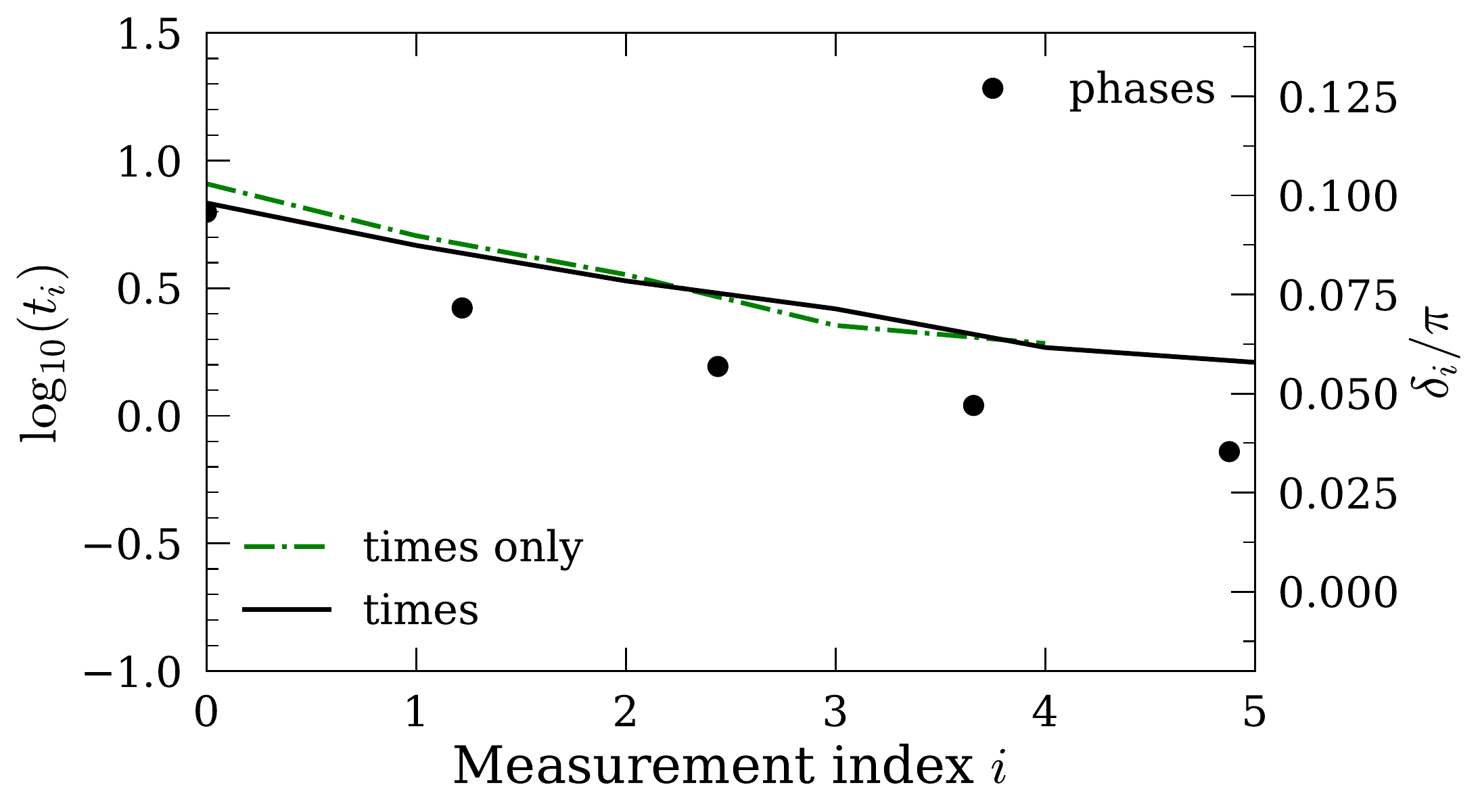}
\caption{Times and phases obtained by the optimization
procedure for the filter function shown in Fig. 1 of the main text. The horizontal axis is 
the measurement index.}
\label{fig:timesandphases}
\end{figure}

In figure \ref{fig:timesandphases} we plot the optimized
times and  phases for this case.  The times are nearly
logarithmically distributed, while the phases are
all positive and larger for the longer times.
The times obtained in an optimization with all phases
set to zero are also shown. Additional
information on the spectra of the state could lead
to different times and phases. Note that switching
the order of the times and phases has no impact in the
absence of errors, in the convergence plots in the paper
we apply the shortest time first.

\begin{figure}[!htb]
\includegraphics[scale=0.4]{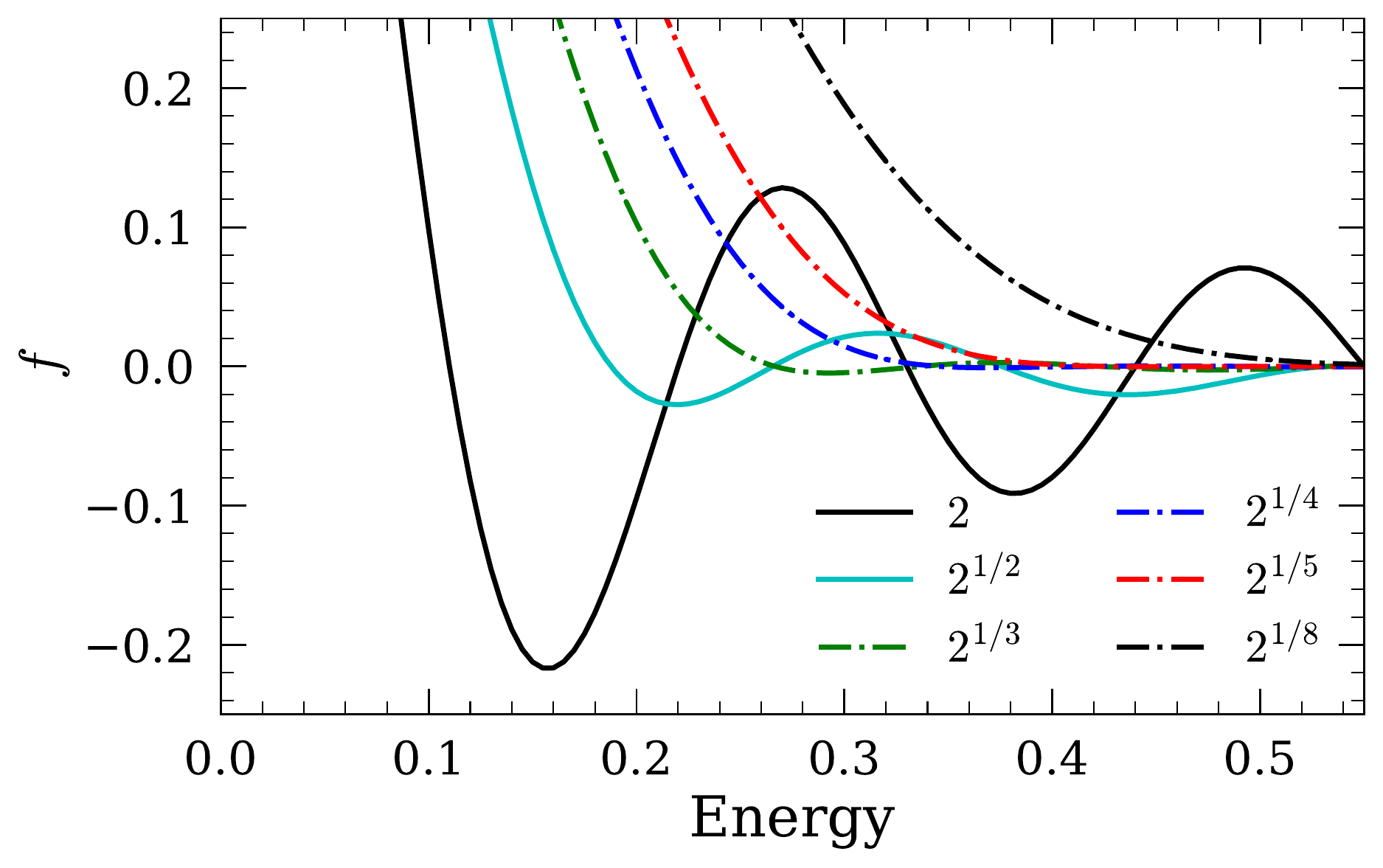}
\caption{ Filter function as a function of energy for a constant total time of $\pi/\Delta$
and exponentially distributed times with various ratios of individual measurement times.}
\label{fig:exponent}
\end{figure}

For large systems the gap will typically be very small compared to the width of the spectrum. This algorithm handles
these cases very well.  When the gap is very small the optimized times approach a constant ratio of $\sqrt{2}$
the filter function for this case has very little ringing,
the filter function for exponential times for various ratios are shown in Fig. \ref{fig:exponent}.

\begin{figure}[!htb]
\includegraphics[scale=0.4]{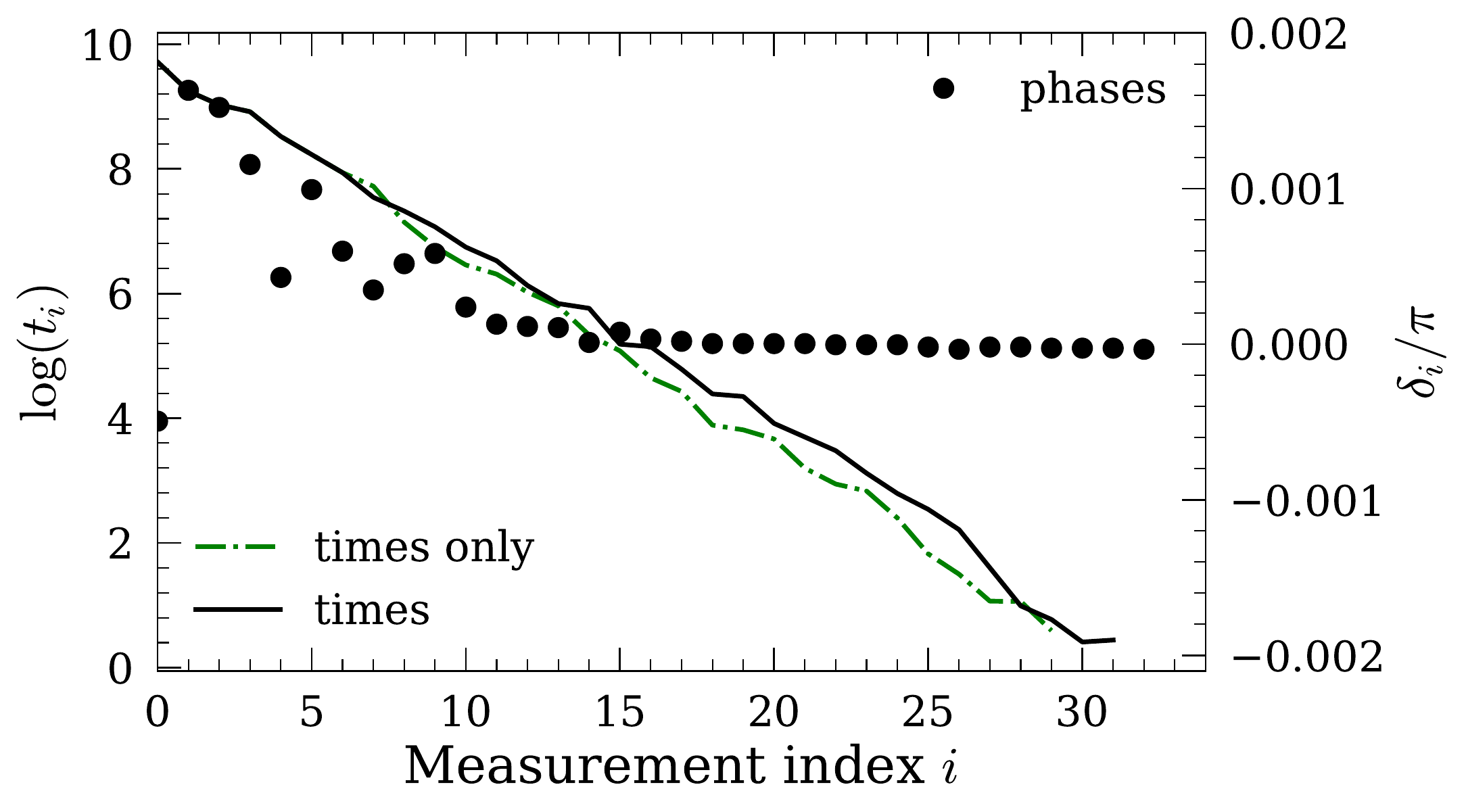}
\caption{ Optimized times and times plus phases for the case of a gap of 0.0001 and a total time of $2 \pi / \Delta$}
\label{fig:times-and-phases-tinygap}
\end{figure}

For a finite gap the optimized times and phases perform slightly better. In Fig. \ref{fig:times-and-phases-tinygap}
we show the optimized times and phases for a total time of $2 \pi/\Delta$  and a gap of 10$^{-4}$. We choose the longer total time to demonstrate the rapid convergence with even longer times. 
The optimized times are nearly exponential for this case. 
The phases for the shorter time step
are indisinguishable from zero in the optimization.

\begin{figure}[!htb]
\includegraphics[width=3in]{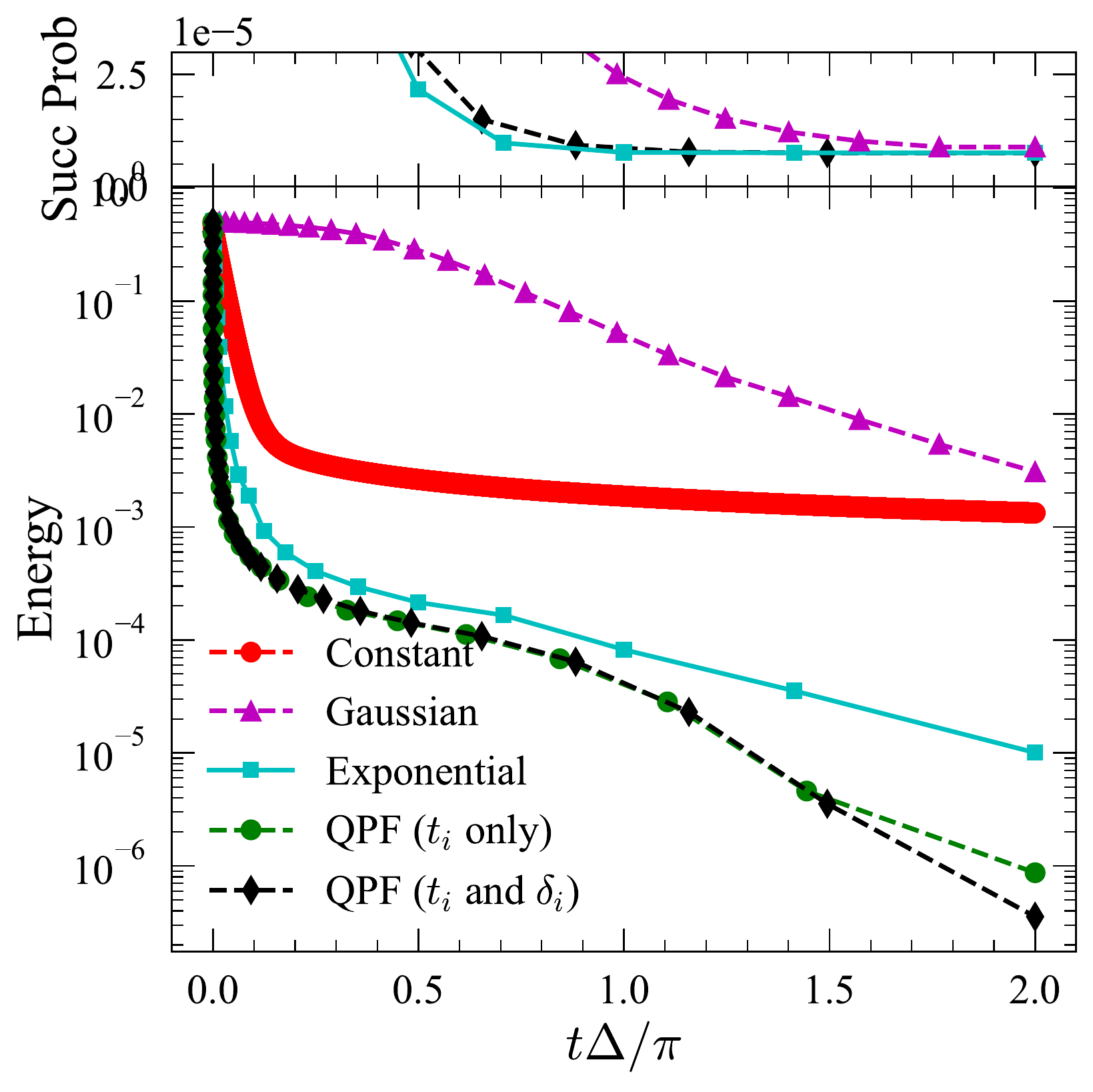}
\caption{ Energy versus total propagation time for a gap of $10^{-4}$ to a time of $2 \pi/\Delta$. }
\label{fig:evst-tinygap}
\end{figure}

The energy versus time for this case is shown
in Fig. \ref{fig:evst-tinygap}. The Hamiltonian eigenvalues are randomly distributed between the gap and one, and the initial state is very poor, random amplitudes on each state. For this reason
the acceptance probability is artificially low.
Nevertheless the convergence is excellent.

\subsection{Applications to the nuclear many-body problem}

\begin{figure}
    \centering
    \includegraphics[scale=0.46]{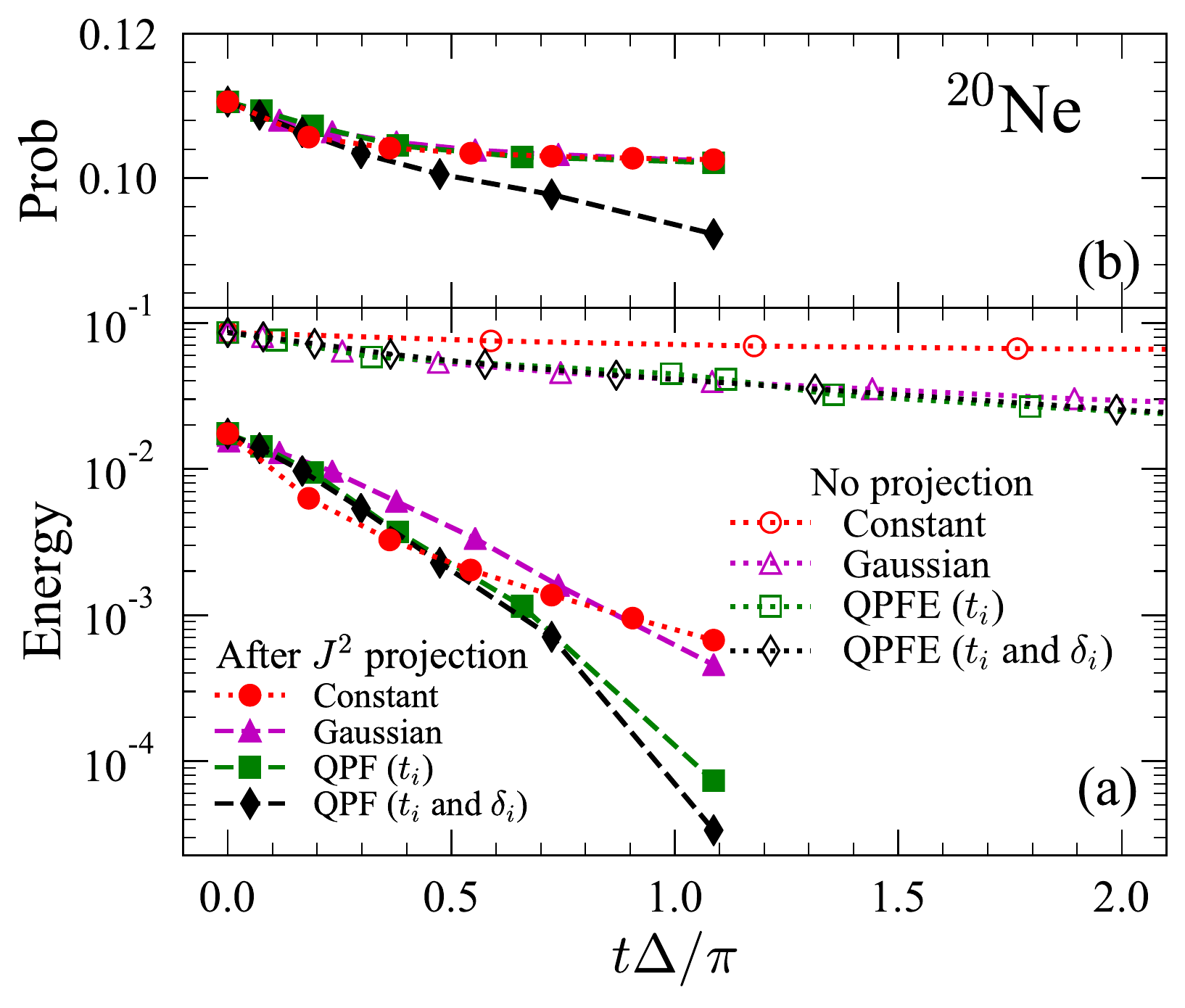}
    \caption{Energy (a) and success probabilities (b) as a function of time for preparing the ground state for $^{20}$Ne in the $sd$ shell model space using the "universal USD" interaction \cite{WILDENTHAL19845,*PhysRevC.74.034315}, starting from the HF solution. We show constant time (red circles), Gaussian-sampled times (purple triangles), QPFE (green squares), and QPFE with optimized times and phases (black diamonds), and with the same filed symbols the Gaussian-sampled times, QPF and QPF with optimized times and phases, but after the trial HF state has been projected on $J=0$. Gaussian sampling produce both positive and negative values, but since the phase is not important in this case, we plot the running sum of absolute values of times. The exact targeted energy is at zero.}
    \label{fig:smapplication}
\end{figure}

\begin{figure}[t]
    \includegraphics[scale=0.46]{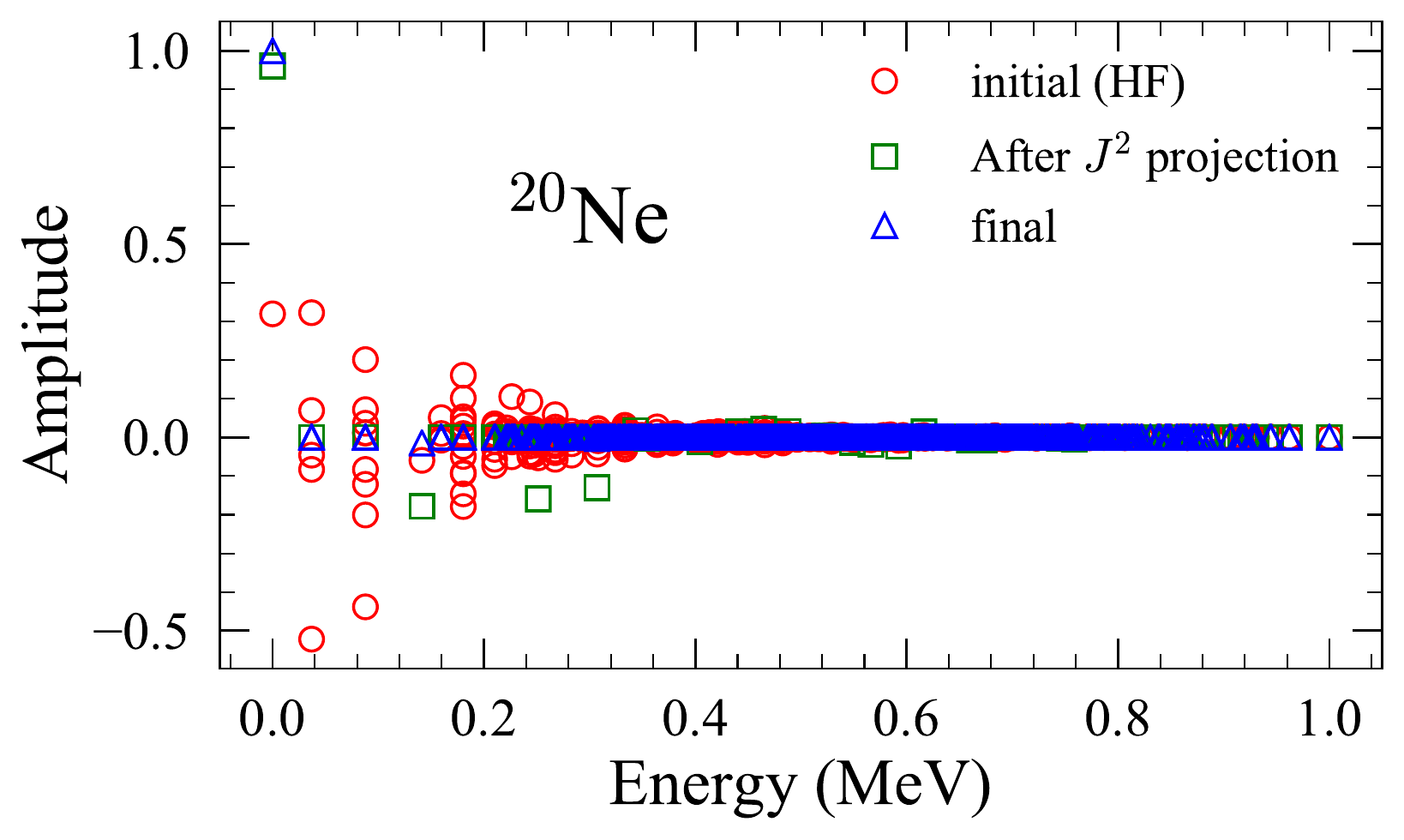}
     \caption{Amplitudes of exact eigenvectors in the decomposition of the HF state (red circles), the state after the $J^2$ projection (green squares), and the final state (blue triangles) for $^{20}$Ne in $sd$ shell.}
    \label{fig:Ne20-amplitudes}
\end{figure}

In general, even-even nuclei have a more regular spectrum ($J=0$ spins for the ground state, larger well-defined gaps, and somewhat clear rotational and vibrational states/energy bands) than their odd-odd and odd-mass counterparts. However, as we have shown in the main text, the algorithm works well for odd-odd nuclei as well. In Fig. \ref{fig:smapplication}, we show the results of different filtering methods for $^{20}$Ne (2 protons and 2 neutrons interacting in the $sd$ shell, with a $^{16}$O core) using the "universal sd" interaction \cite{WILDENTHAL19845,*PhysRevC.74.034315}. In this case, the ground state has $J=0$ spin, so the operator applied in Eq. (1) of the main text is $O=J^2$. We also show in Fig. \ref{fig:Ne20-amplitudes} the same amplitudes for the problem considered in Fig. \ref{fig:smapplication}. The HF state contains more of the higher states, in particular first and second excited states, which are degenerated (given their values of the total angular momentum). After spin projection, however, the $J=0$ in this case are the only surviving ones, while further energy filtering removes the excited $J=0$ states.

\begin{figure}[t]
    \includegraphics[scale=0.46]{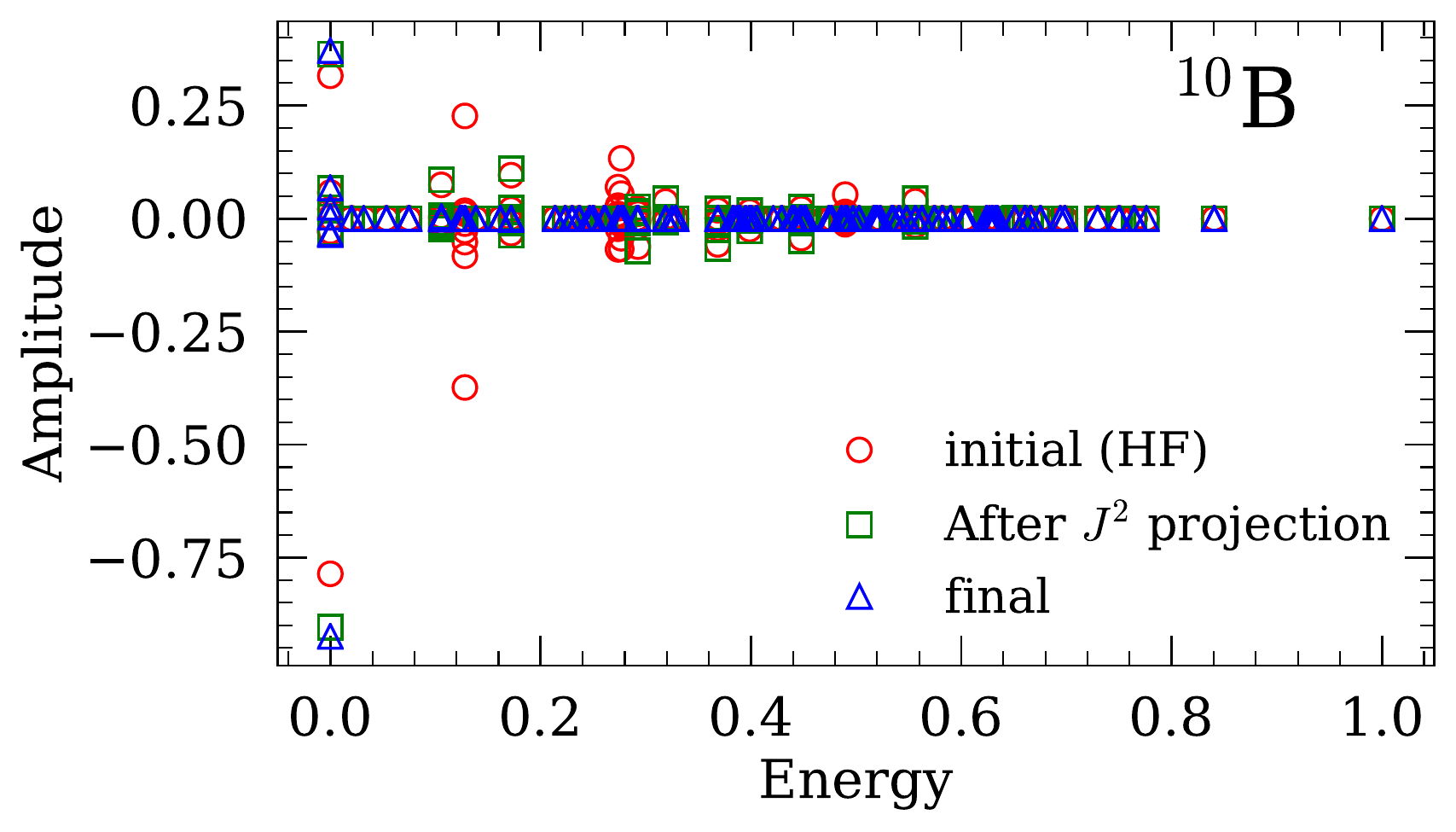}
    \caption{Same as in Fig. \ref{fig:Ne20-amplitudes} for $^{10}$B corresponding to the example presented in Fig. 3 of the main text. Because the state is degenerated, it is a mixture of different spin projections. Hence probability of those states add to one.}
    \label{fig:B10-amplitudes}
\end{figure}

In Fig. \ref{fig:B10-amplitudes} we show additional results for the $^{10}$B case presented in Fig. 3 of the main text. Here we plot the amplitudes of the exact energies in the Hartree-Fock (HF) state, after the $J^2$ projection, and at the end, after we apply the QPF with optimized times and phases. Note that the final state we obtain is a superposition of seven-degenerated spin projection values, and this is why we do not obtain a single blue triangle with amplitude 1.

\begin{figure}
    \centering
    \includegraphics[scale=0.46]{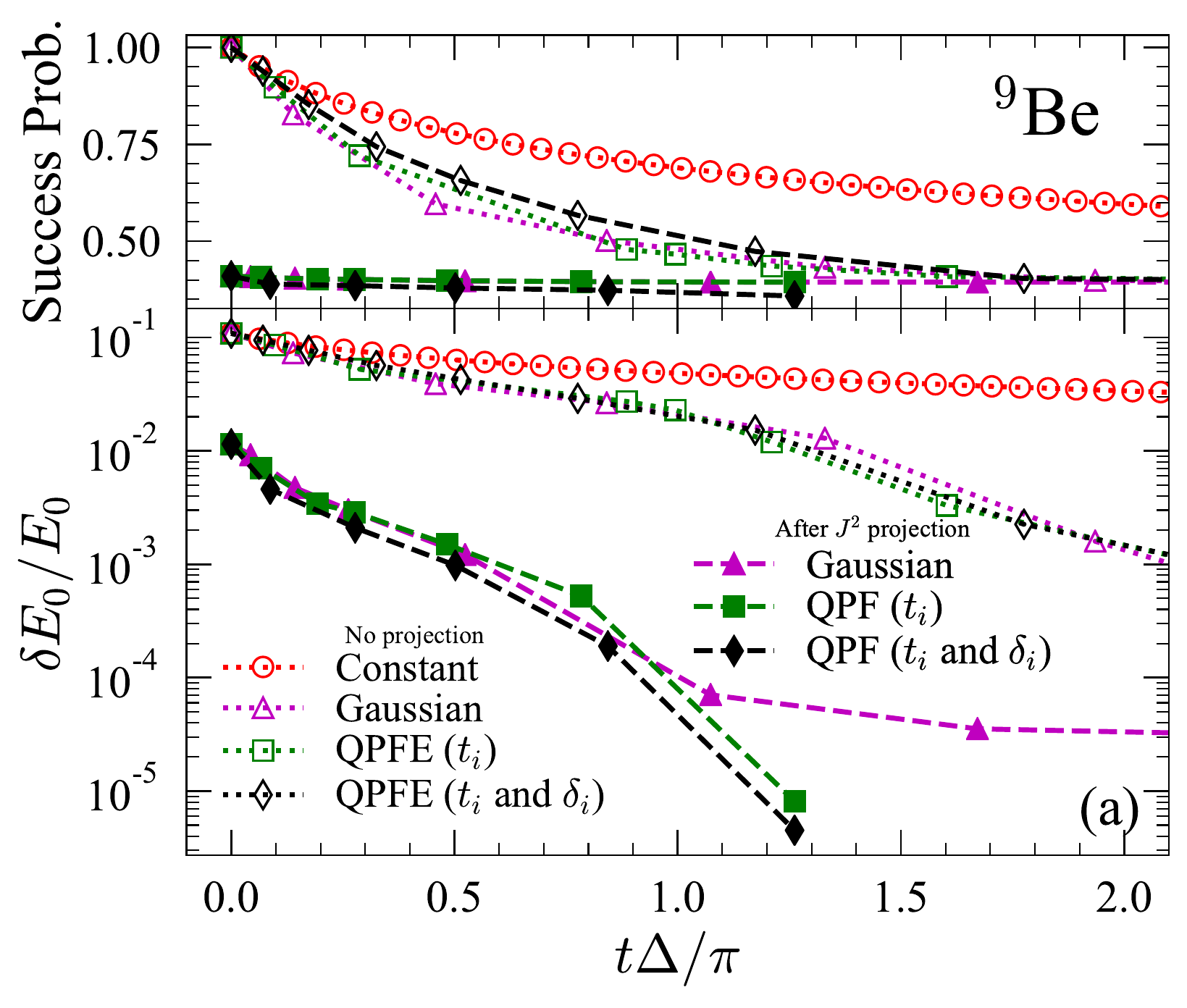}
    \caption{Relative error in energy (a) and success probabilities (b) as a function of time for preparing the ground state for $^{9}$Be (2 protons and 3 neutrons in $p$ shell). In this case, the targeted has $J=3/2$ spin, hence the operator used to project on good spin is $J^2-3.75$.}
    \label{fig:b9}
\end{figure}

Finally, in Fig. \ref{fig:b9}, we also present an odd-even nucleus, $^9$B in the p-shell, with the Hamiltonian given by the Cohen-Kurath interaction \cite{COHEN19651} in the $p$ shell. In this case, since the ground state is spin $3/2$, for the $J$ projection we have shifted $J^2$ by $3.75$. As shown in the figure, even in this case, our algorithm also provides most improvement. 

\subsection{Success Probabilities for the general algorithm}
We report here a description of the repeated measurement scheme.
Given a state $\ket{\Psi}$, and ancillary qubit in state $\ket{0}$ and a family of unitary operators $U_i$ acting as
\begin{eqnarray}
U_i\ket{0}\ket{\Psi}&=&\ket{0}A_i\ket{\Psi}+\ket{1}B_i\ket{\Psi}
\end{eqnarray}
where $A_i$ and $B_i$ are general hermitian operator.
After evolution with $U_1$ operator and a projective measurement over the ancillary register we have the state
\begin{eqnarray}
\ket{\psi_1}=\frac{\ket{0}A_1\ket{\Psi}}{\sqrt{p_1}}\, ,
\end{eqnarray}
with $p_1=\bra{\Psi}A_1^2\ket{\Psi}$ success probability for obtaining the outcome $0$ over the ancillary qubit. Repeating the scheme above, evolving with an operator $U_2$, whose action over the state $\ket{\psi_1}$ is described below
\begin{eqnarray}
U_2\ket{0}\ket{\psi_1}&=&\frac{\ket{0}A_2 A_1\ket{\Psi}}{\sqrt{p_1}}+\frac{\ket{1}B_2A_1\ket{\Psi}}{\sqrt{p_1}} \, ,
\end{eqnarray}
and after measurement postselection of the ancilla in state $0$ we obtain
\begin{eqnarray}
    \ket{\psi_2}=\frac{\ket{0}A_1A_2\ket{\Psi}}{\sqrt{p_2}}
\end{eqnarray}
where
\begin{eqnarray}
p_2=\bra{\Psi}A_2^2A_1^2\ket{\Psi}\, .
\end{eqnarray}
and we notice the cancellation of $\sqrt{p_1}$.  Generalizing the above identities can be generalized in a straightforwad way to the case of N consecutive measurement leading to the following success probability
\begin{eqnarray}
p_N=\bra{\Psi}A_N^2\cdots A_2^2A_1^2\ket{\Psi}\, .
\end{eqnarray}
If we consider a set of operators of the form $A_i=f_i(H-E)$, where $f_i$ some general block encoding of a function of the Hamiltonian we can write
\begin{eqnarray}
p_N&=&\bra{\Psi}\prod_if^2_i(H-E)\ket{\Psi}\\
&=&\sum_n\lvert\bra{\Psi}n\rangle\rvert^2 \prod_i f^2_i(E_n-E)\, ,
\end{eqnarray}
and choosing the set of functions $f_i$ peaked near the ground state energy $E_0$ and vanishing for the remaining energies we can write
\begin{eqnarray}
p_N\simeq \lvert\bra{\Psi}0\rangle\rvert^2 \prod_i f^2_i(E_0-E)\, ,
\end{eqnarray}
which is the formula reported in the main text.

\bibliography{references}